\documentclass[reprint,pre,floats,floatfix,aps,amsmath,amssymb]{revtex4-2}
\usepackage{graphicx}
\begin{document}
\title{Thinning algorithms for the Monte Carlo simulation
	of kinetic Ising models}
\author{V. I. Tokar}
\affiliation{G. V. Kurdyumov Institute for Metal Physics of the
N.A.S.\ of Ukraine, 36 Acad. Vernadsky Boulevard, UA-03142 Kyiv, Ukraine}
\author{H. Dreyss\'e}
\affiliation{Universit{\'e} de Strasbourg, CNRS, IPCMS, UMR 7504,
F-67000 Strasbourg, France}
\date{\today}
\begin{abstract}
The thinning method for numerical generation of the nonhomogeneous Poisson
process (NHPP) arrival times has been adapted to accelerate Monte Carlo
simulations of the kinetic Ising models (KIMs) with the Glauber spin-flip
dynamics. The performance of the suggested algorithms has been illustrated
by  simulation of the decay of metastable states in stationary KIMs and
of hysteresis in KIMs in a periodic external field. The thinning has been
implemented by means of piecewise constant majorizing functions which
exceed or are equal to NHPP rate. It has been shown that in favorable
cases the use of thinning makes possible the simulations of hysteresis
at frequencies in tens nanohertz and the decay of metastable states with
lifetimes by many orders exceeding those in previous simulations. Good
agreement of simulated results with low-temperature analytic theories
has been established. Though the  algorithm acceleration has been shown
to enhance with lowering temperature, the hysteresis has been simulated
at moderately low temperatures of practical interest estimated to cover
the range from below the room temperature up to temperatures used in
hyperthermia applications.
\end{abstract}
\maketitle
\section{Introduction}
The kinetic Ising models (KIMs) have been widely used in modeling
diverse systems in many fields \cite{interdisciplinary24}.  Besides
traditional use in studies of uniaxial magnets, lattice gases, and alloy
kinetics \cite{ducastelle}, the KIM framework has been also adapted
to study protein folding \cite{BPZ}, optimization of neural networks
\cite{aguilera_unifying_2021,Kirkpatrick1983OptimizationBS}, and many
other topics \cite{interdisciplinary24}.

A major reason for KIMs widespread use is their convenience for simulation
by the kinetic Monte Carlo (KMC) method \cite{landau2009guide,KMCreview07}
which can provide a reliable quantitative insight into the
model behavior. However, in many cases the computer time required
for simulation becomes prohibitively long which severely reduces
the practicality of KMC. The problem is usually exacerbated at low
temperatures where, on the one hand, long-lived metastable states tend
to occur \cite{binder-nucleation}, on the other hand, the performance
of the conventional Metropolis algorithm (MA) decreases due to small
acceptance rates \cite{metropolis,n-fold,landau2009guide}.

To address the problem, KMC algorithms have been
developed that make possible considerable acceleration
of KMC simulations for KIMs with stationary Hamiltonians
\cite{n-fold,novotny1995,KMC1999,trygubenko,opplestrup2006,PRE08,%
KMC1999,KMCreview07}.
However, some important phenomena can be studied only with
Hamiltonians that explicitly depend on time. A well known
example is the hysteresis in oscillating magnetic field---the
phenomenon which is not only of considerable scientific interest
but also has important technical and biomedical applications
\cite{chakrabarti_dynamic_1999,Novotny-nanomagnets2007,hyperthermia}.

In the absence of more efficient techniques, hysteresis in KIMs has been
studied with the use of MA \cite{2DIsing1998,zhu_hysteresis_2004}. Because
of its limitations, simulations were performed either at a high
temperature in order to enhance the acceptance rate, or at relatively
large frequencies when more cycles of the external field can be simulated
during available computer time.

However, in biomedical applications the temperature in 80\% of the
Curie temperature used in KMC simulations of Ref.\ \cite{2DIsing1998}
is too high in the case of iron and magnetite nanoparticles often used in
hyperthermia studies \cite{hyperthermia} because their Curie temperatures
are of order of $ 10^3 $~K. The simulations at lower temperatures in
Ref.\ \cite{zhu_hysteresis_2004}, on the other hand, did not cover the
frequency range of interest in biomedical studies \cite{hyperthermia}.

A detailed discussion of physical behaviors
encountered in KIM simulations can be found in Refs.\
\cite{rikvold1994,novotny1995,2DIsing1998,droplet-theory98,%
novotny2002-3D,zhu_hysteresis_2004}.
In the present study the main focus will be on KMC performance.
Our aim is to suggest accelerated algorithms which would make possible
KMC simulation in magnetic nanoparticles in temperature and frequency
ranges of practical interest. This will be achieved by adapting to the
non-stationary case the Bortz, Kalos, and Lebowitz algorithm (BKLA) that
enhances KMC performance at low temperatures \cite{n-fold} and the MC
algorithm with absorbing Markov chains (MCAMC) or the Novotny algorithm
(NA) \cite{novotny1995} which proved to be effective in simulating the
decay of metastable states \cite{novotny1995,novotny-review2002}.

Formally, the difference between KMC simulation of the
stationary and non-stationary processes is that in the
latter case the transition rates of the stochastic matrix are
time-dependent. Therefore, one is faced with a necessity to simulate
a non-stationary or non-homogeneous Poisson process (NHPP). This
problem has been extensively covered in literature (see, e.g., Refs.\
\cite{streit2010poisson,nhpp-wiley,ross2010introduction,piecewise_nhpp,%
cpd-theorem})
and a general conclusion was that, apart from some exceptions, in general
case the fastest algorithms to generate NHPP can be obtained by the
thinning approach \cite{thinning} with the use of piecewise-constant
majorizing function. By the latter is meant a function which is larger
or equal to the rate it majorizes. We adopt this terminology from Ref.\
\cite{nhpp-wiley} though in mathematical literature a different definition
of majorization is more common \cite{majorization}.

In this study we have applied thinning algorithms to simulations of
KIM of a ferromagnet in an oscillating magnetic field at temperatures
and frequencies of practical interest and to the decay of metastable
states in the stationary KIM. We will prepend letter ``T'' to their
abbreviations (TBKLA and TNA) to distinguish our algorithms from their
counterparts. But it should be remembered that due to the freedom of
choice of the majorizing function, thinning can be implemented in a
variety of ways.  We will develop two algorithms,---TBKLA and TNA,---based
on a constant majorization function and TNA with a piecewise constant
majorizing rate.  To justify quantitative comparison with simulations
based on MA \cite{2DIsing1998,zhu_hysteresis_2004} we will show that
the conventional MA is equivalent to a thinning algorithm which means
that the physical time simulated by MA is the same as in our algorithms.

The paper is organized as follows. In the next section we concretize
our KIM. In Sec.\ \ref{thinning} a simple case of thinning with a
constant majorization will be applied to BKLA and the resulting TBKLA
will be used to illustrate its advantage over MA at low temperatures
in a high-frequency hysteresis simulation. In Sec.\ \ref{NA} NA will
be accelerated with the use of thinning and  in Sec.\ \ref{lifetime}
the efficiency of TNA will be illustrated by the simulations of decay of
metastable states. In Sec.\ \ref{periodic} TNA will be further extended
by the use of piecewise-constant majorization to simulate KIM in a
periodic external field; in Sec.\ \ref{hysteresis} the low-frequency
behavior of hysteresis loop areas will be simulated at two temperatures
and at the lower temperature good agreement with the adiabatic theory
at low frequencies will be demonstrated. In the concluding section a
brief summary and possibilities of further improvement of the developed
algorithms will be discussed.
\section{The model}
To simplify simulations and to facilitate
comparison with theoretical predictions of Refs.\
\cite{expEbeta1991,rikvold1994,SD-rates,nita2003} and with previous KMC
simulations \cite{novotny1995,2DIsing1998,novotny2002-3D}, we consider
ferromagnetic KIMs with nearest neighbor (nn) spin interactions. The
spins occupy  $ N $ sites of a toroidal hyper-cubic lattice with $
L $ lattice sites in each of $ D $ directions ($ D=2,3 $), so $ N=L^D
$. In this geometry all spins are equivalent which greatly simplifies
combinatorics. In real nanostructures the boundary spins would differ
from those in the bulk, but generalization to this more realistic case
should be straightforward, though more computationally demanding.

We will deal with the Ising Hamiltonian 
\begin{equation}\label{hamiltonian}
	{\cal H} = -J\sum_{\langle i,j\rangle}s_is_j - H(t)\sum_{i} s_i,
\end{equation}
where  $ J >0$ is the pair spin interaction and $ s_i=\pm1 $ are the
Ising spins. The first summation on the right hand side (rhs) is over
all nn spin pairs and the second one over $ N $ lattice sites. $ H(t)
$ is the oscillating external magnetic field of the form
\begin{equation}\label{Ht}
	H(t)=-H_0\cos(\omega t)
\end{equation}
with $ H_0 >0$ being the field amplitude and $ \omega=2\pi\nu $, where $
\nu $ is the oscillation frequency; the stationary case is recovered at
$ \omega=0$. More general periodic fields, e.g., those studied in Ref.\
\cite{2componentH}, can be treated similarly.

In the Glauber dynamics the spin reversal (or flip) rate reads 
 \cite{glauber} 
\begin{equation}\label{rates}
r_i(t) =	\frac{\nu_0}{1+\exp\left(\beta\Delta_i{\cal H}\right)},
\end{equation} 
where $ \nu_0 $ is a characteristic frequency in the range $ 10^9-10^{12}
$~Hz \cite{2DIsing1998}; $ \beta=1/T $ where temperature $ T $ is in
energy units which is achieved by setting the Boltzmann constant to unity
\cite{novotny2002-3D,soft-dynamics,zhu_hysteresis_2004}; the Hamiltonian
change under spin $ i $ reversal $ s_i\to-s_i $ in nn KIM is
\begin{equation}\label{DH}
	\Delta_i{\cal H} = -2Js_i\sum_{nn}s_{i_{nn}}+2H_0s_i\cos(\omega t)
\end{equation} 
where $ s_i $ is the spin value after the flip and  $ s_{i_{nn}} $
are its nearest neighbors.  The total spin flip rate in the system
\begin{equation}\label{lambda-t}
	\lambda(t)=\sum_{i}r_i(t)
\end{equation}
is the NHPP rate which have to be simulated. 

Formally the problem is as follows. Given previous arrival time $ t_0 $,
by using Eqs.\ (\ref{cpd}) and (\ref{f=u}) in Appendix \ref{appa} it is
easy to show that the next arrival $ t $ can be found as the solution
of equation
\begin{equation}\label{the-eq}
\int_{t_0}^{t}\lambda(t^\prime)dt^\prime=-\ln u,
\end{equation}
where $ u $ is a random variate uniformly distributed within the range
$ 0< u<1 $; besides, the use has been made of the fact that $ 1-u $
and $ u $ are distributed identically.  Here a terminology remark
is in order. Throughout the paper we will use term ``arrival time''
or simply ``arrival'' for the value of the evolution parameter as
generated by a Poisson process (PP) both homogeneous or nonhomogeneous
\cite{streit2010poisson}. In our opinion, the alternative term ``event''
\cite{nhpp-wiley} can be misleading in the thinning approach that
we are going to use because nothing happens if the generated time
is rejected. However, each arrival generated with the exact Eq.\
(\ref{the-eq}) does lead to an event, to a spin flip in the case of the
Glauber dynamics.

From Eq.\ (\ref{lambda-t}) it is seen that in general a numerical
solution of Eq.\ (\ref{the-eq}) can be computationally costly because
in iterative solution methods the quadrature will have to be computed
numerically many times.

However, in some cases the problem simplifies. For example, if $
\lambda $ does not depend on time, the solution is easily found as
\cite{landau2009guide}
\begin{equation}\label{const-lambda}
	t= t_0-\ln u/\lambda.
\end{equation}
Somewhat more complex but still computationally easy case
provides a piecewise-constant rate when $ \lambda(t) $
remains constant on consecutive time intervals with the time
dependence restricted to discontinuities at the interval boundaries
\cite{ross2010introduction,piecewise_nhpp,nhpp-wiley,streit2010poisson}.
These particular cases in conjunction with the thinning technique allow
for a fast generation of NHPP with the rates as complex as $ \lambda(t)
$ Eq.\ (\ref{lambda-t}).
\section{\label{thinning}The thinning algorithm}
The KMC simulation of NHPP with the use of thinning algorithms is based
on the theorem of Ref.\ \cite{thinning} which states that any NHPP
with rate function $ \lambda(t) $ can be generated through thinning
of arrival times generated with the use of a majorizing rate function
$ \bar{\lambda}(t)\geq \lambda(t) $ as follows (see, e.g., Ref.\
\cite{streit2010poisson}). The algorithms proceed as follows. First
one generates an increasing sequence of arrival times $ \{t_i\} $, $
i=1,2,\dots $ within interval $ t_0< t_i\leq t_{max} $ ($ t_{max} $
is an arbitrarily chosen maximum time) using Eq.\ (\ref{the-eq}) with $
\bar{\lambda}(t) $ instead of $ \lambda(t) $ as
\begin{equation}\label{bar-the-eq}
	\int_{t_{i-1}}^{t_{i}}\bar{\lambda}(t)dt=-\ln u^\prime_i, 
\end{equation} 
where $ u^\prime_i $ are independent uniform random variates. Next a new
sequence of the variates $ \{u_i\} $ is generated and for each $ t_i $
the acceptance criterion
\begin{equation}\label{criterion}
	\bar{\lambda}(t_i) u_i \leq \lambda(t_i)
\end{equation}
is tested. If the arrival time $t_i$ does not satisfy the criterion
it is discarded while the remaining ``thinned'' set  $\{t_i\}^{th}$
gives a statistically the same sequence of the arrival times as if
generated with the use of Eq.\ (\ref{the-eq}) \cite{thinning}. By using
a time-independent (constant) majorizing function the sequence $ \{t_i\}
$ can be easily generated with the use of Eq.\ (\ref{const-lambda}) and
because the acceptance criterion Eq.\ (\ref{criterion}) does not include
quadratures the generation of $\{t_i\}^{th}$ that we are interested in
becomes faster than with the use of Eq.\ (\ref{the-eq}) with nontrivial $
\lambda(t) $.

To proceed further, we first note the difference between simulation
setups usually discussed in mathematical literature  on NHPP
\cite{thinning,ross2010introduction,piecewise_nhpp,nhpp-wiley,streit2010poisson}
and the KIM simulations. In the former case one considers rate $
\lambda(t) $ as an immutable function and therefore the batch simulation
of arrivals is warranted. But in KIMs the rate is known only for an actual
spin configuration $  e $ acquired by the system at earlier arrival $
t_e $ which should be updated at every configuration change. In fact,
in the case of KIMs the rate should be more comprehensively written as $
\lambda(t,e,t_e) $, but because this is always the case, we will omit
two last arguments for brevity. Thus, only the nearest arrival time is
reasonable to generate to avoid superfluous computations.

Therefore, in using the thinning method we will test the acceptance
criterion at every arrival according to the following general thinning
algorithm (GTA) that will be used throughout the paper.  For definiteness,
the simulation time span was chosen to be from $ t=0 $ to $ t_{max} $. The
initial configuration $ e_0 $ at $ t_e=0$ can in principle be  arbitrary,
but in our low temperature simulations we found convenient to chose it
as the fully ordered state with negative magnetization. Explicit forms of
the corresponding $ \bar{\lambda}(t) $ and $ {\lambda}(t) $ at $ t_e=0 $
should be obtained with the use of appropriate definitions, such as Eq.\
(\ref{lambda-t}).  Then the following three steps should be repeated
until termination:
\begin{description}
	\item [Step 1] Set $ t_0=t_e$,	generate uniform random variate $
	u\in(0,1) $, and solve Eq.\ (\ref{bar-the-eq}) for the arrival
	time $ t $ by setting $ t $ as the upper integration limit, $
	t_0 $ as the lower one, and $ u $ as the logarithm argument;
	if $ t > t_{max} $ terminate.
	\item [Step 2] Calculate $ \lambda(t) $ and generate random
	variate $ u\in(0,1) $; if $u\bar{\lambda}(t)>\lambda(t) $ return
	to \textbf{Step 1}.
	\item [Step 3] For $u\bar{\lambda}(t)\leq\lambda(t) $ set $ t_e
	=t $ and determine new configuration $ e $ as stipulated by the
	KIM dynamics; define $ \bar{\lambda}(t) $ and $ {\lambda}(t) $
	corresponding to $ e $ and $ t_e$; return to \textbf{Step 1}.
\end{description}
 
We note that there would be no need to differentiate between $ t_0 $ and $
t_e $ if the exact Eq.\ (\ref{the-eq}) was used in the simulation because
in this case all arrivals would cause the configuration change. But the
distinction is important in the presence of rejected arrivals because
they do not change the system and so cannot influence NHPP rates which
may depend on $ t_e $.
\subsection{Equivalence of MA to a thinning algorithm}
Arguably, MA is the most widely known MC algorithm. In view of
differing definitions, we will understand by MA the one called in
Ref.\ \cite{n-fold} the standard MC algorithm. Though initially
developed for the stationary case \cite{metropolis}, MA has
been widely used in the hysteresis simulations (see, e.g., Refs.\
\cite{2DIsing1998,zhu_hysteresis_2004}). Therefore, our aim is to show
that MA is equivalent to a thinning algorithm of the KIM that we are
dealing with so the physical results obtained with the use of MA should
be the same as in our simulations.

Let us consider a constant majorizing rate 
\begin{equation}\label{lambda0}
	\bar{\lambda}=N\nu_0 
\end{equation} 
which  exceeds $ \lambda(t) $ in Eq.\ (\ref{lambda-t}) for any values
of the parameters, as can be seen from Eq.\ (\ref{rates}).

\begin{figure}
	\includegraphics[scale = 0.75]{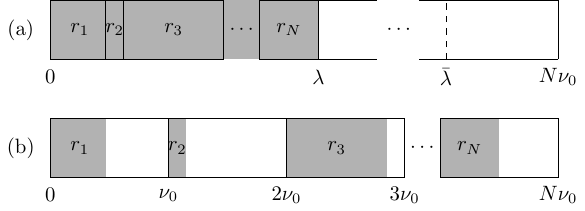}
	\caption{\label{fig1}Visualization of interrelation between BKLA
	with rate $ \lambda $ \cite{n-fold}, conventional MA with total
	rate $N\nu_0 $, improved MA with $ \bar{\lambda}=N\bar{r} $ Eq.\
	(\ref{bar-r}), and the thinning.  Shaded areas correspond to
	accepted arrivals, empty space to rejected ones; for detailed
	explanation see the text.}
\end{figure}

Assuming that current time is $ t_0 $ and using majorizing constant $
\bar{\lambda} $ as $ \bar{\lambda}(t)$ we may generate $ t $ using Eq.\
(\ref{const-lambda}). In our GTA we now have to calculate $ \lambda(t)
$ using Eq.\ (\ref{lambda-t}) and to test the acceptance criterion by
generating a uniform random variate $ u $. This is visualized in Fig.\
\ref{fig1}(a) where the test reduces to the question about whether $
u\lambda=uN\nu_0 $ falls into the empty region of the rectangle or in
the shaded one.

An advantage of MA is that there is no need to calculate $ \lambda(t)
$ in Eq.\ (\ref{lambda-t}) by accounting for all $ r_i $ in Eq.\
(\ref{rates}) which may be computationally costly, for example, in the
case of long range interactions. Instead, one first randomly chooses one
of the spins,say, $ i $, and then compares $ u\nu_0 $ with only $ r_i(t)
$. In Fig.\ \ref{fig1}(b) the acceptance or rejection would correspond to
whether $ u\nu_0 $ falls into the shaded or open region within cell $ i $.

The statistical equivalence of this procedure to GTA is a consequence
of the uniform distribution property that the probability of $ uN\nu_0
$ falling into, say, shaded region is the same in both rectangles in
Fig.\ (\ref{fig1}). The number of the spin that need be tested can
be found as $ i=[uN]+1 $, where $ [\cdots] $ means the integer part,
and acceptance/rejection is decided from the fulfillment/violation
of inequality $ uN-[uN]=\{uN\}\leq r_i(t) $ ($ \cdots$ denotes the
fractional part).

Thus, MA simulates the same KMC process as GTA, though there is a small
difference. In practice usually in MA a constant time increment $ \Delta t
= t-t_0 = 1/\nu_0N $ \cite{metropolis} is used instead of the stochastic
value from Eq.\ (\ref{const-lambda}) \cite{KMC1999}. Though the latter
approach is more rigorous, if one is interested only in an average
evolution obtained in as many simulation runs as possible, the use of
constant increment can be fully justified for stationary Hamiltonians. The
random variates in the increment simulation and for determining the
spin flips are statistically independent, so the time increment can
be averaged analytically which amounts to $ \langle\Delta t\rangle =
1/\nu_0N $. However, in nonstationary case the spin flip rate will depend
on the value of time increment and thus correlate with it. However, for
the time variation in Eqs.\ (\ref{hamiltonian})--(\ref{Ht}) on the scale
much larger than $ \Delta t $ the constant increment approximation should
quite accurate. Condition  $ \omega\Delta t \ll1 $ can be rewritten as
\begin{equation}\label{nuLLnu0}
	\nu \ll N\nu_0
\end{equation}
so with $ \nu_0 $ is in the range $ 10^9-10^{12} $~Hz \cite{2DIsing1998}
the condition in practical applications may be violated only in extreme
cases \cite{KMC1999}.

The advantages of MA are that there is no need in calculating  $ r_i(t)
$ for all $ i $ and $ \bar{\lambda} $  at each KMC step; the only needed
quantity is $ \bar{r}=\nu_0 $ known beforehand. Moreover, to accelerate
MA it can be lowered by using instead
\begin{equation}\label{bar-r}
	\bar{r}=\max_{e,i,t} r_i(t)
\end{equation}
which is always smaller than $ \nu_0 $, as can be seen from Eq.\
(\ref{rates}). The average time step in this case is $ \langle\Delta
t\rangle =1/N\bar{r}$. This can be especially useful at high temperatures
where $ \bar{r} $ approaches $ \nu_0/2 $ \cite{n-fold,KMC1999} making
the use of MA at such temperatures more efficient.

However, among all configurations $ e $ in Eq.\ (\ref{bar-r}) there
always exist configurations with spin $ i $ being in an energetically
unfavorable state which at low temperatures would lead to $ \bar{r}\approx
\nu_0 $. But in a fully ordered state the spin reversal rates at low
temperatures are very small for all $ i $ which under MA would cause
many rejections resulting in a slow evolution \cite{n-fold}.

In contrast, the sum of all rates $ r_i $ in Eq.\ (\ref{lambda-t}) in
the ordered state will be small for moderate values of $ N $, so with
the use of an appropriate $ \bar{\lambda} $ the simulation might proceed
faster despite the extra computations needed.  This can be achieved
with the use of BKLA technique \cite{n-fold} where the computation of $
\lambda(t) $ in Eq.\ (\ref{lambda-t}) is greatly facilitated in nn KIM.
\subsection{\label{tbkla}Simulation of high-frequency hysteresis using TBKLA}
As temperature lowers the spin flips become less frequent. But the time
step in MA remains the same---$ 1/N\nu_0 $ which means that the rejection
rate grows. Below some temperature there will be so many rejections
that the calculation of the rejected individual spin rates $ r_i $ would
eventually exceed the computational cost of updating the total rate Eq.\
(\ref{lambda-t}) at each KMC step. This happens especially fast in the
case of nn spin interactions when only a small number of rates have to be
updated \cite{n-fold}.  For example, in the case of hyper-cubic lattice
each spin has only $ 2D $ nearest neighbors.  In the stationary case
$ 2(2D+1) $ different spin flip rates should be calculated and stored
only once. Though updating the transition rates at each MC step requires
additional computations, the absence of rejections in BKLA makes it more
efficient than MA already at moderately low temperatures \cite{n-fold}.

To generalize BKLA on the nonstationary case by means of thinning it is
possible in principle to make the rejection rate in TBKLA arbitrarily
small by using sufficiently tight majorization function. However, to
reduce the computational overhead a compromise should be sought between
the  admissible rejection rate and the complexity of $ \bar{\lambda}
$. Besides, the simplicity of implementation would facilitate the
adaptation of TBKLA to possible model modifications.

Therefore, similar to MA, let us consider the simplest case
of majorization by a constant rate. This can be achieved by
using $ \bar{\lambda} $ of the same form as $ \lambda(t) $ in Eq.\
(\ref{lambda-t}) but with all rates $ r_i(t) $ replaced by their maximum
values
\begin{equation}\label{bar-rates}
	\bar{r}_i=\max_tr_i(t) =	\frac{\nu_0}{1+\exp\left(\beta\bar{\Delta}_i{\cal H}\right)}
\end{equation}
where
\begin{equation}\label{DbarH}
	\bar{\Delta}_i{\cal H} = \min_t {\Delta}_i{\cal H}= -2Js_i\sum_{nn}s_{i_{nn}}-2 H_0.
\end{equation}
Here used has been made of Eq.\ (\ref{DH}) and of our choice $
H_0\geq0 $.  An advantage of the constant majorization is that the $
2(2D+1) $ different majorizing rates $ \bar{r}_i $ can be calculated
and stored only once, though the set of independent $ r_i(t) $ should be
recalculated at each KMC step because of their dependence on time. The
computational overhead will depend on the rejection rate.

As can be seen from Eqs.\ (\ref{bar-rates}) and (\ref{DbarH}), when both
$ \beta $ and $ H_0 $  are sufficiently large  $\bar{r}_i\simeq \nu_0
$ for all spins. In other words, we essentially recover MA which is
inefficient at low temperatures. Thus, for the constant majorization to
be efficient the external field amplitude should be sufficiently small
for the argument of the exponential function in Eq.\ (\ref{bar-rates}),
that is, rhs of Eq.\ (\ref{DbarH}), was positive. This means that the
amplitude should satisfy inequality
\begin{equation}\label{Hc}
	H_0 < 2DJ =H_{DSP}(T=0)
\end{equation} 
where $ H_{DSP}(T)  $ is the  value of the external field at the dynamic
spinodal point \cite{rikvold1994}. But even if $ H_0 $ satisfies the
inequality, TBKLA performance would worsen when $ H_0 $ would be too
close to $ H_{DSP}(0) $.
\subsubsection{Asymmetric hysteresis}
\begin{figure}
	\includegraphics[bb=148 464 394 668,scale = 0.9]{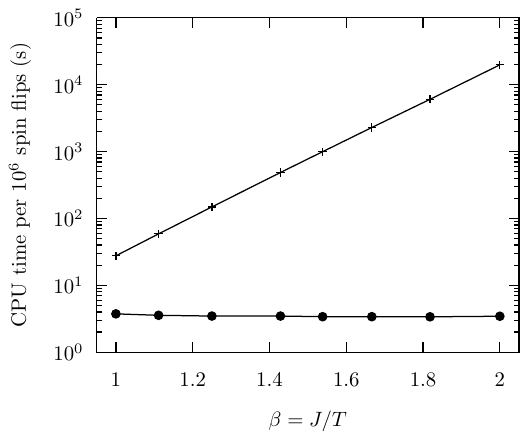}
	\caption{\label{fig2}Temperature dependence of execution time
	of MA (crosses) and TBKLA (diamonds) in simulations of $ 10^6 $
	spin flips in 2D KIM of size $ L=100 $ in oscillating magnetic
	field of frequency $ \omega=\nu_0 $ and amplitude $ H_0=J $.}
\end{figure}

To illustrate the efficiency of TBKLA we compared it with simulations by
MA of asymmetric hysteresis in KIM of Ref.\ \cite{zhu_hysteresis_2004}. In
our implementation of MA both the constant and the random time
increments have been used but no do difference of the results have
been noticed. The results presented in Fig.\ \ref{fig2} show that, as
expected, the efficiency of TBKLA with respect to MA improves with the
lowering temperature. For example, at $ \beta=2/J $ BKLA has been found
to be almost four orders of magnitude ($ 6\cdot10^3 $ times) faster
than MA.  The computer time in the figure corresponds to processor
AMD Ryzen 5 4500U in a virtual environment, so we estimate the base
frequency to be $\gtrsim2$~GHz.  This processor was used in all our
simulations. Its characteristics can be relevant because according to
Ref.\ \cite{piecewise_nhpp} the performance of thinning algorithms may
crucially depend on microprocessor architecture.
The efficiency of TBKLA for simulation of the asymmetric high-frequency
hysteresis to a large degree is a consequence of the fact that
the evolution takes place in the vicinity of stable states which
at low temperatures are almost fully ordered (see Fig.\ 2 in Ref.\
\cite{zhu_hysteresis_2004}).  Both $ \lambda(t) $ and $ \bar{\lambda} $
in this case are dominated by the smallest individual spin rates which
makes the time steps in Eq.\ (\ref{const-lambda}) large.
\subsubsection{Symmetric hysteresis}
From standpoint of practical applications, such as hyperthermia,
more interesting is the symmetric hysteresis due to much larger loop
areas \cite{hyperthermia}. The symmetric loops occur at sufficiently
low frequencies \cite{DPT2017} so we tested TBKLA in this case by
simulating 2D KIM of size $ L=64 $ \cite{2DIsing1998} using two sets of
parameters. One set was $ T=J $ and $ H_0=3.5J $ and another one $ T=0.5J
$ and $ H_0=1.5J $. Both amplitudes $ H_0 $ satisfy Eq.\ (\ref{Hc})
but in the first case due to closeness to $ H_{DSP}(0) $ and higher
temperature the system is in the deterministic regime.  This has been
established by simulating the average nucleation time $ \langle\tau\rangle
$ as described in Refs.\ \cite{rikvold1994,novotny2002-3D}. It has been
found that it does not depend on system size and is $ \simeq\nu_0^{-1}
$ which means that there was no energy barrier to spin reversals. The
average hysteresis curves remained symmetric at all simulated frequencies
up to $ \nu\leq100\nu_0 $.

\begin{figure}
	\includegraphics[bb=148 463 486 671,scale=0.7]{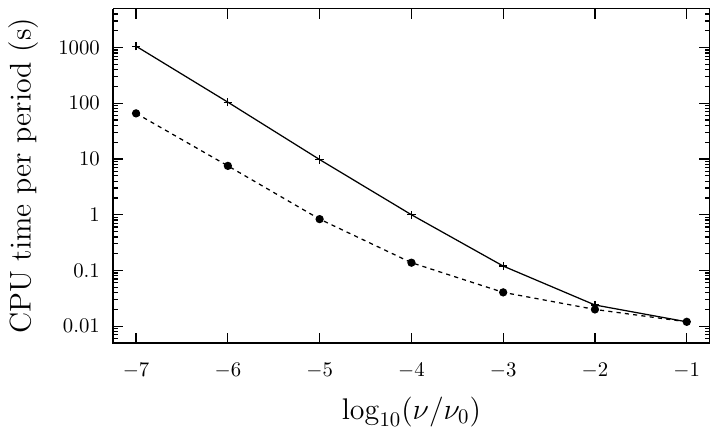}
	\caption{\label{fig3}Dependence of simulation time on the
	frequency of oscillations of the external field for 2D KIM of size
	$ L=64 $ with $ H=3.5J $ at temperature $ T=J $. Crosses: TBKLA;
	bullets: TNA of Secs.\ \ref{periodic} and \ref{hysteresis}.}
		\end{figure}

\begin{figure}
	\includegraphics[bb=148 463 488 671,scale=0.69]{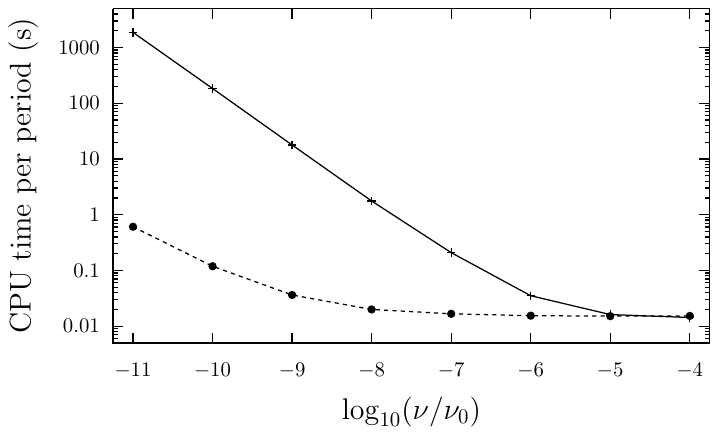}
	\caption{\label{fig4} Same as in Fig.\ \ref{fig3} for $ H=1.5J $ and $ T=0.5J $.}
		\end{figure}

The performance of TBKLA is illustrated by upper graphs in Figs.\
\ref{fig3} and \ref{fig4}. As can be seen, the algorithm
makes possible in a reasonably short computer time simulate hysteresis
in the frequency region $ 10^{5}-10^{6} $~Hz often used in hyperthermia
even for $ \nu_0$ as large as $10^{12} $~Hz (using estimates of $
\nu_0\sim10^9-10^{2} $~Hz of Ref.\ \cite{2DIsing1998}). As expected,
at lower temperature and $ H_0 $ further from $  H_{DSP}(0)$
the much longer periods at the same computer time could be
simulated. However, still smaller frequencies can be of experimental
interest \cite{hyst-thin-films,Fe_on_W}. Besides, faster simulations
would made possible gathering larger statistics. Therefore, further
acceleration of KMC algorithms at low frequencies would be desirable.
As can be seen in Figs.\ \ref{fig3} and \ref{fig4}, at low
frequencies the simulation time is proportional to the oscillation period
$ \bar{t}=1/\nu $ which can be easily understood  qualitatively. According
to Eqs.\ (\ref{hamiltonian})-(\ref{lambda-t}), the time variation of $
\lambda(t) $ becomes weak at small $ \omega $. And because the average
rates at all frequencies are the same (this will be explicitly shown
in the next section), the KMC steps  on average are of the same size at
small frequencies so the simulation time should be proportional to the
period length. Thus, because on average the longer the period the more
KMC steps are needed to cover it, the algorithm acceleration should be
sought in reducing the number of simulated KMC steps.
\section{\label{NA}Thinning the Novotny algorithm}
One way of reducing the number of KMC steps in thinning algorithms is
to use a tighter majorization, e.g., including more steps into piecewise
constant functions. This could have improved the performance at moderately
low frequencies shown in Figs.\ \ref{fig3} and \ref{fig4}.
However, even in stationary systems when BKLA is fully rejectionless,
the simulation of the magnetization switching in metastable states
encounters a severe slowdown due to the low rate of nucleation of the
opposite spin domains \cite{rikvold1994}. A solution to this problem
was suggested in Ref.\ \cite{novotny1995} using the observation that
at low temperatures the KMC evolution is plagued by the back and forth
fluctuations between a small number $ s $ of transient states close
to the fully ordered one without advancing the growth of nucleation
droplets. But the linear master equation with  $ s\times s $ matrix
of transition rates governing the fluctuations between a small number
of states can be integrated faster, even analytically in some cases,
than simulated by KMC methods. When out of the subspace the simulation
proceeds by the standard BKLA \cite{novotny1995}. We will call this
intermittent use of BKLA and MCAMC the Novotny algorithm (NA).

In simulations of hysteresis this approach can be even more useful
because the system spends most of the time near one of two stable free
energy minima where the fluctuations are completely unproductive because
the total magnetization reversal is virtually impossible in even small
nanostructures.

In the present study we will adapt an appropriately modified NA formalism to hysteresis simulation as follows. First, we will use stochastic matrices in continuous time instead of the discrete-time Markov chains. Second, we will deal with non-stationary Hamiltonians. Finally, to accelerate the algorithm we will use thinning and will call it TNA. For simplicity, its implementation will be illustrated on the simplest case $ s=2 $ but generalization to larger $ s $ should be feasible along similar lines. 

Before proceeding, a remark about our notation. Because ferromagnetic
Ising model has two fully ordered states with positive and negative
magnetizations, our notation $ e_0 $ is ambiguous so it would be natural
to distinguish the states, e.g., by using a superscript as $ e_0^\pm $.
However, we will need many quantities to describe evolution of states $
e_0-e_s $ which all would have to carry the superscript. So for brevity we
will develop formalism for one of the ordered states which can be modified
to the use in the other state via a simple reversal of spin directions.

In $ s=2 $ case we have only two states $ e_0 $ and $ e_1 $ which
satisfy continuous-time evolution equation with $ 2\times2 $ matrix of
non-stationary rates  $ R(t) $
\begin{equation}\label{dot_psi}
	\frac{d\psi_k(t,t_0)}{dt}=R(t)\psi_k(t,t_0),
\end{equation}
where $ \psi_k $, $ k=0,1 $ are two independent 2-component vectors
\begin{eqnarray}\label{psi0}
	\psi_0(t,t_0)&=&\begin{bmatrix}
		p_1(t,t_0)\\ 1-p_1(t,t_0)
	\end{bmatrix}\\ \psi_1(t,t_0)&=&\begin{bmatrix}
	1-p_0(t,t_0)\\ p_0(t,t_0)
\end{bmatrix}.
\end{eqnarray}
The evolution is initiated by $ p_k(t_0,t_0)=0 $ so that vectors $
\psi_k(t_0,t_0) $ can be identified with states $ e_k $ $ k=0,1 $; $
p_1 $ is the probability of finding reversed spin when the evolution
starts at $ e_0 $ and similarly for $ p_0 $. Explicitly,
\begin{equation}\label{R}
	R(t)=\begin{bmatrix}
		-r_{\underline{q}}(t) & Nr_{\underline{0}}(t)\\
		r_{\underline{q}}(t) & -Nr_{\underline{0}}(t)
	\end{bmatrix}, 
\end{equation} 
where underlined subscripts are the rates Eq.\ (\ref{rates}) which have
the underlined number of nn spins pointing in the direction opposite to
spin $ i $ and $ q=2D $ is the coordination number.

Our generalization of the algorithm of Refs.\ \cite{novotny1995} proceeds
as follows. If at time $ t_e $ the system state is $ e_0 $, further
evolution will follow the $ e_0\leftrightharpoons e_1 $ fluctuations
described by Eq.\ (\ref{dot_psi}) for $\psi_0(t,t_e) $ until the second
spin is reversed in which case following Ref.\ \cite{novotny1995} the
simulation proceeds via TBKLA until a new occurrence of state $ e_0
$. Because in PP two spins cannot reverse simultaneously, the two-spin
state can appear only when at least one reversed spin is already present
in the system. Thus, the second reversal is given by the product of the
probability of having one reversed spin $ p_1(t,t_{e_0}) $ multiplied
by the rate of the second spin reversal in state $ e_1 $
\begin{equation}\label{lambdatt0}
	\lambda(t)=p_1(t,t_{e_0})r_{1\to2}(t)
\end{equation}
where \cite{novotny2002-3D}
\begin{equation}\label{r12}
	r_{1\to2}=\left[(N-q-1)r_{\underline{0}}+qr_{\underline{1}}\right]
\end{equation}
is the rate of the second spin reversal in the presence of one reversed
spin.  Though only  $p_1(t,t_{e_0}) $  is needed in TNA, the problem is
that it depends on two arguments which means that a two-parameter surface
would have to be calculated and/or stored  which can be a computationally
demanding task. To facilitate it, we invoke the transition matrix
formalism \cite{brockett1970} as explained in Appendix \ref{transition}.
\section{\label{lifetime}Lifetime of metastable states}
To illustrate superior performance of TNA over MCAMC let us
consider the decay of metastable states in the stationary KIM
\cite{SD-rates,novotny2002-3D,Shneidman_2003,nita2003,Shneidman_2005}. In
this case all $ r_i $ are time-independent so integrals in
Eqs.\ (\ref{p_it}) and (\ref{p_1tt0}) are easily taken to give
\begin{equation}\label{p_1}
	p_1(t,t_0)=\mu^{-1}Nr_{\underline{0}}\left[1-e^{-\mu(t-t_{e_0})}\right],
\end{equation}
Where $ \mu $ as defined in Eq.\ (\ref{mu}) is time-independent in the stationary case.
Substituting this in Eq.\ (\ref{lambdatt0}) one finds
\begin{eqnarray}\label{2ndspin}
	&&\lambda(t)=p_1(t,t_{e_0})\left[(N-q-1)r_{\underline{0}}+qr_{\underline{1}}\right]\nonumber\\
	&&=\frac{r_{\underline{0}}}{\mu}N\left[(N-q-1)r_{\underline{0}}+qr_{\underline{1}}\right]\left[1-e^{-\mu(t-t_{e_0})}\right].
\end{eqnarray}
As can be seen, by omitting the negative term in the second brackets we can obtain a constant majorization for $ \lambda(t) $ as
\begin{equation}\label{major}
	\bar{\lambda}=\frac{r_{\underline{0}}}{\mu}N\left[(N-q-1)r_{\underline{0}}+qr_{\underline{1}}\right].
\end{equation}
Moreover, the inequality $ \bar{\lambda} \geq \lambda(t) $ is quite
tight because the neglected term quickly turns to zero as $ t-t_{e_0}$
grows.  By assuming that $ N $ is fixed,  it is easy to see that as $
T\to0 $ $ \bar{\lambda} \to0$ because both rates in the nominator $
r_{\underline{0}} $ and $ r_{\underline{1}}  $ vanish in this limit while
$ \mu\to1 $ as can be seen from Eq.\ (\ref{mu})) where the dominant rate
$ r_{\underline{q}}\to1 $.  Thus, at sufficiently small temperatures
the random time difference generated with the use of $ \bar{\lambda}$
in Eq.\ (\ref{const-lambda}) will be on average much larger than $
\mu^{-1}\simeq1 $ so the term with minus sign in Eq.\ (\ref{2ndspin})
will be negligible. This means that in a typical time step at low enough
temperature the rejections will be very rare.

For comparison purposes we have applied TNA (in conjunction with
conventional BKLA because of the stationary Hamiltonian) to determine
the nucleation times $ \langle\tau\rangle $ in two metastable
systems simulated in Ref.\ \cite{novotny2002-3D} within NA. Also,
KMC simulations have been compared with theoretical predictions of
Refs.\	\cite{expEbeta1991,novotny2002-3D,nita2003} according to which
the nucleation time is equal to the mean time of escape over an energy
barrier. In the case of single-domain (SD) nucleation the latter depends
on the barrier height $ \Gamma_D $ as 
\begin{equation}\label{tau}
	\langle\tau\rangle=\frac{A_D}{\nu_0N}\exp\frac{\Gamma_D}{T},
\end{equation}
where dependence of $ A_D $ on $ T $ is weaker than in the Arrhenius
law and at low temperatures is negligible. The prefactor $ 1/N $ comes
from the fact that SD nucleation rate which is the inverse of $ \tau $
is proportional to the number of nucleation sites in the system $ N $.

\begin{figure}
	\includegraphics[bb=148 464 492 668,scale=0.7]{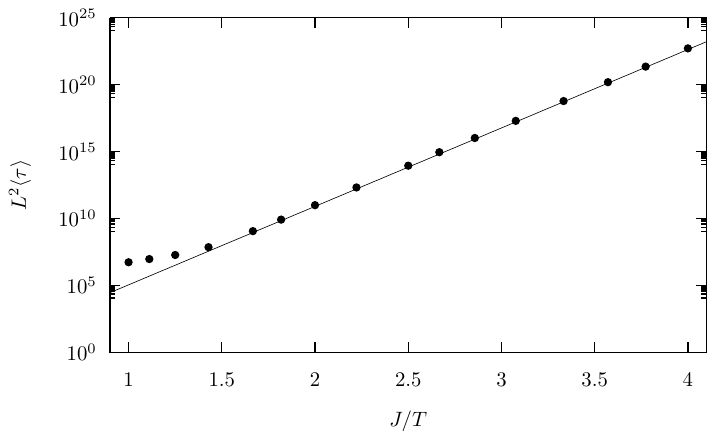}
	\caption{\label{fig5}Mean lifetimes of 2D KIM of size $ L=200 $
	with $ H=3J/4 $ as in Fig.\ 1 of Ref.\ \cite{novotny2002-3D}
	simulated with the use of TNA (symbols); straight line
	is the low-temperature theoretical prediction Eqs.\
	(\ref{rate-t})-(\ref{gamma}).}
\end{figure}

The first simulated system was 2D KIM that have the largest size $
L=200 $ and the highest nucleation barrier among 2D models studied in
long-time simulations in Ref.\ \cite{novotny2002-3D} within $ s=3 $ MCAMC
algorithm (NA in our terminology). Though we have used only $ s=2 $ TNA,
we succeeded in simulating almost 6 order of magnitude longer nucleation
times, as can be seen from comparison of our Fig.\ \ref{fig5} with
Fig.\ 1 of Ref.\cite{novotny2002-3D}. The maximum average CPU time of
simulation for one nucleation at $ T=0.25J $ was less than 1.5 min which
means that even longer times could have been simulated. But agreement
with theoretical prediction seen in Fig.\ \ref{fig5} is already very
good so interpolation to lower temperatures by Eq.\ (\ref{tau}) should
be reliable.

Another example was a $ 3D $ KIM  with $  H_0= 3.9J$ studied within
MCAMC algorithm in Ref.\ \cite{novotny2002-3D} where the small-$
T $ interpolation was not quite straightforward (see Fig.\ 4 in
that reference). This particular case is of interest because in Ref.\
\cite{novotny2002-3D} the zero-temperature limit of the measured barrier
height
\begin{equation}\label{G3T}
	\Gamma_3(T)= T\ln(L^3 \langle \tau\rangle)
\end{equation}
was predicted to be
\begin{equation}\label{G30}
	\Gamma_3(0)=28J-6H_0=4.6J
\end{equation}
in strong disagreement with  $\Gamma_3(0) = J $ predicted in other
studies. However, the lowest temperature $ T=0.09J $ in simulations
of Ref.\ \cite{novotny2002-3D} was rather high and in view of large
statistical errors the interpolated value at $ T=0 $ was not determined
precisely despite the use of  up to $ s=4 $ MCAMC algorithms.

\begin{figure}
	\includegraphics[bb=170 468 412 673,scale = 0.80]{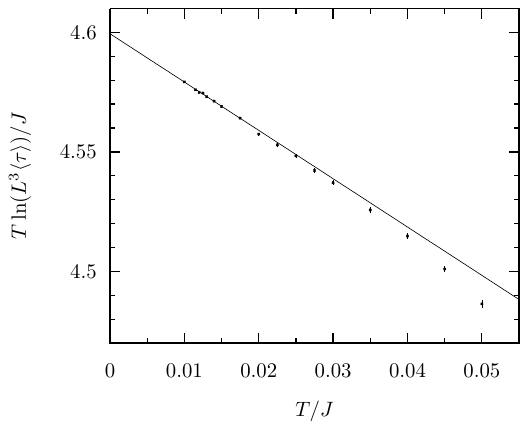}
	\caption{\label{fig6}Simulation of nucleation from metastable
	state with the use of $ s=2 $ TNA in simple cubic system with $
	L=32 $ and $\vert H_0\vert=3.9J $ \cite{novotny2002-3D} at low
	temperatures. The number of nucleations $ N_{nucl}$ at each $ T $
	was $10^3$; the function at the vertical axis is the rhs of Eq.\
	(\ref{G3T}). Linear interpolation (thin solid line) has been
	done over the points with $ T\leq0.015J $.}
\end{figure}

In Fig.\ \ref{fig6} are shown our results of $ s=2 $ TNA simulations
extended down to $ T=0.01J $. As is seen, an excellent agreement with
Eq.\ (\ref{G30}) has been found. The average time of simulation for
one nucleation event grew from 0.5~s at the highest temperature in
the figure to $ \sim10 $~min at $ T=0.01J $. Because the nucleation
events are completely independent, statistics can be improved with
the use of parallel architecture, so the simulations could have been
performed even at smaller temperatures. Furthermore, because of the
form of Eq.\ (\ref{G3T}), according to the error propagation formula
the statistical errors scale with the number of nucleation events $
N_{nucl} $ as $ T/\sqrt{N_{nucl}} $. Therefore, at $ T=0.01J $ the same
accuracy as at $ 0.1J$ could have been achieved with 100 times smaller $
N_{nucl} $. However, in Fig.\ \ref{fig6}  statistics was gathered over $
N_{nucl} =10^3$ at all temperatures  in order to assess the algorithm
performance. It may be expected that even faster simulations are possible
with the use of  $ s>2 $ algorithms \cite{novotny1995,novotny-review2002}.

Thus, our simulations have shown that TNA is more efficient than
MCAMC which we attribute to the fact that in MCAMC one effectively
deals with solving a discrete analogue of Eq.\ (\ref{the-eq}) which is
computationally more time consuming in comparison with the thinning
method.
\section{\label{periodic}TNA for periodic field}
If one is interested in using TNA to study long-time behavior of
nonstationary KIM with an arbitrary time dependence of $ H(t) $, a
problem arises of computing and storing transition matrices on large
time intervals. A periodicity of the external field makes possible
to deal with the transition matrix known only inside one oscillation
period. This is achieved with the use of the Floquet's theorem  according
to which the transition matrix can be cast in the form  (see, e.g.,
Ref.\ \cite{brockett1970})
\begin{equation}\label{floquet}
	P(t,t_0)=S^{-1}(t)e^{K(t-t_0)}S(t_0),
\end{equation}
where $ S $ and $ K $ are, respectively, a periodic and a constant
matrices. Thus,  it would be sufficient to store matrix elements of $
S $ only within one period $ 0\leq t<\bar{t} $. However, if a high
accuracy is needed the performance of TNA may depend on the processor
memory architecture.

To simplify the simulations, in the present study we have developed a
computationally easy approximate approach for low-frequency hysteresis
because, as can be seen in Figs.\ \ref{fig3} and \ref{fig4},
the exact and  simple in implementation TBKLA performs quite well at
higher frequencies. At low frequencies $ \omega/\nu_0 $ can be used as a
small parameter to develop simple and computationally efficient formalism,
as will be shown below.

To facilitate numerical estimates it will be convenient to switch to
dimensionless quantities
\begin{eqnarray}\label{x}
&&	x = \omega t, \\
\label{2}
&&\omega=\omega/\nu_0, \; t=\nu_0 t, \mbox{\ and}\\
\label{3}
&&\lambda(x)=\lambda(t)/\nu_0.
\end{eqnarray}
This should not cause confusion because definitions on the second line
can be viewed as simply choosing $ \nu_0 $ as the frequency unit and when
using $ \lambda(x) $ Eq.\ (\ref{3}) it should be remembered that $ x $
is defined in Eq.\ (\ref{x}). The use of $ x $ as the evolution parameter
simplifies hysteresis analysis because now the oscillation period is
always $ 2\pi $ irrespective of the frequency. In this notation Eq.\
(\ref{the-eq}) becomes
\begin{equation}\label{the-eq4}
	\int_{x_0}^{x}\lambda(x^\prime)dx^\prime = -\omega\ln u.
\end{equation}
As is seen, the left hand side of this equation does not depend on $
\omega $ so $ \Delta x = x-x_0=O( \lambda) $ which shows that the number
of KMC steps needed to cover period $ 2\pi $ should be proportional to
$ 2\pi/\omega=1/\nu =\bar{t}$, as has been pointed out in the previous
section.

To compute $ \lambda $ Eq.\ (\ref{lambdatt0}) we need to calculate
probability $ p_1 $ Eq.\ (\ref{p_1tt0}) which in new notation reads
\begin{equation}\label{p_1xx0}
	p_1(x,x_{e_0})=\left[\hat{p}_1(x)-\chi(x,x_{e_0})\hat{p}_1(x_{e_0})\right],
\end{equation} 
where
\begin{equation}\label{change}
	\hat{p}_1(x)=\omega^{-1}\int_{0}^{x}\exp\left[-\omega^{-1}\int_{x^\prime}^{x}\mu(x^{\prime\prime})
	dx^{\prime\prime}\right]a_1(x^\prime)dx^\prime
\end{equation}
(see Eq.\ (\ref{p_it})) and
\begin{equation}\label{exp}
	\chi(x,x_{e_0})=\exp\left[-\omega^{-1}\int_{x_{e_0}}^{x}\mu(x^\prime)dx^\prime\right].
\end{equation}
Because even in a thinning approach we will need $ \lambda(x) $ Eq.\
(\ref{lambdatt0}) to test the acceptance criterion Eq.\ (\ref{criterion}),
there still will be need to compute integrals in Eqs.\ (\ref{change})
and (\ref{exp}), though only once for each arrival. Because frequent
numerical integrations may considerably slow down the simulation, a
better approach might be to use a suitable interpolation over a grid of
points where values of $ \lambda $ can be stored in advance. However,
as we mentioned earlier, to reach simultaneously fast execution speed
and high accuracy, processors with large cache memory may be required.

In our small frequency approximation we avoid numerical integration
in Eq.\ (\ref{change}) by taking the integral analytically using its
Laplace's form. According to Appendix \ref{Laplace} at small $ \omega $
it can be approximated as
\begin{equation}\label{p_1laplace}
	\hat{p}_1(x)=\frac{a_1(x)}{\mu(x)}+O(\omega),
\end{equation}
where higher order terms in $ \omega $ can also be determined
\cite{bender1999advanced} but for simplicity we will neglect $ O(\omega)
$ terms which, in particular, means that our approximations will improve
as frequency lowers.

\begin{figure}
	\includegraphics[bb=148 464 486 667,scale = 0.7]{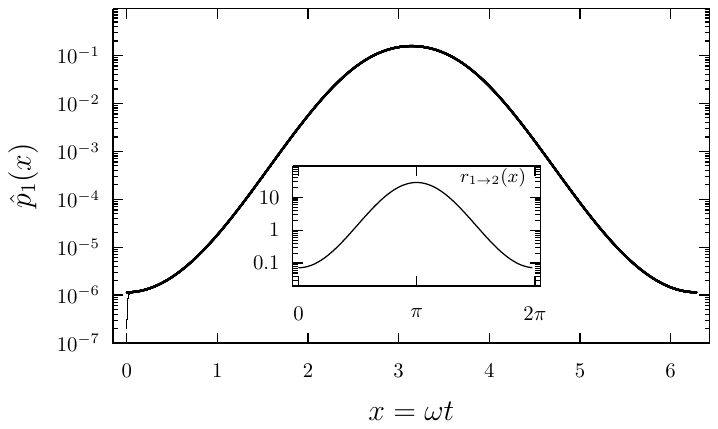}
	\caption{\label{fig7} Solid line: numerical integration in Eq.\
	(\ref{p_it}) compared to the approximate analytic expression Eq.\
	(\ref{p_1laplace}) (dots). The 2D KIM parameters were $ L=64 $,
	$ T=0.5J $, $ H=1.5J $, and $ \omega=10^{-2}\nu_0 $; the inset
	shows rate $ r_{1\to2} $ Eq.\ (\ref{r12}) in units of $ \nu_0 $.}
\end{figure}

\begin{figure}
	\includegraphics[bb=183 406 441 657,scale=0.85]{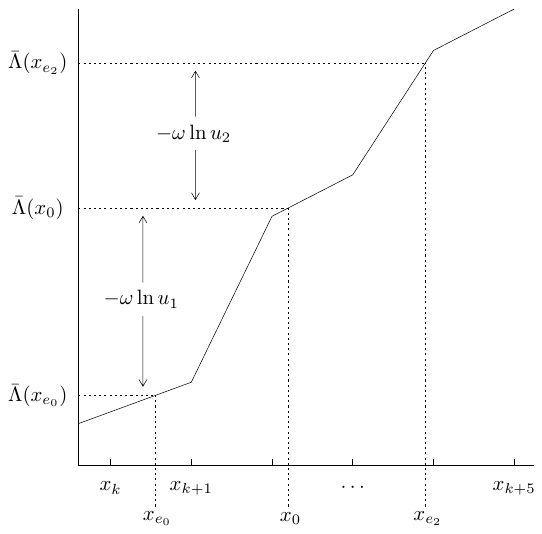}
	\caption{\label{fig8}Schematic illustration of $ s=2 $ TNA
	hysteresis simulation with the use of a piecewise linear
	cumulative majorizing function $ \bar{\Lambda}(x) $ Eq.\
	(\ref{Lambda}). It has been assumed that at point $ x_{e_0} $
	TBKLA simulation reaches a fully ordered state $ e_0 $. Then
	$ \bar{\Lambda}(x_{e_0} ) $ is calculated as described in
	the text and random variate $ u_1 $ is generated. The next
	arrival point $ x_0 $ is determined and another random variate $
	u_1^\prime $ is generated to test the acceptance criterion Eq.\
	(\ref{criterion}). If the arrival is rejected, yet another
	random variate $ u_2 $ is generated and $ x_{e_2} $ is found as
	above. Here subscript $ e_2 $ has been chosen because this time
	we assume that $ u_2^\prime $ satisfies the acceptance criterion
	so the configuration should contain two reversed spins. From
	this point the simulation proceeds by TBKLA.}
\end{figure}

The quality of approximation Eq.\ (\ref{p_1laplace}) has been tested by
a numerical integration over $ 10^4 $ points in Eq.\ (\ref{p_it}) for
the sinusoidal field Eq.\ (\ref{Ht}). From Fig.\ \ref{fig7} it can be
seen that  even at relatively high frequency $ \omega=0.01 $ the worst
discrepancy is restricted to a narrow interval $ \Delta x\sim O(\omega)
$near $ x=0 $ where the exact value should vanish while the approximation
is finite. This means that the errors will be $ O(\omega) $ in agreement
with the chosen level of approximation.  The error introduced by the
discrepancy is further reduced by prefactor $ r_{1\to2} $ in the NHPP
rate Eq.\ (\ref{lambdatt0}), as can be seen in Fig.\ \ref{fig7}.
To estimate the quadrature in Eq.\ (\ref{exp}) we first point out its
difference from similar integration in Eq.\ (\ref{change}). In the latter
case it may cover an arbitrarily large part of the oscillation period $
2\pi $, that is, is an $ O(1) $quantity, while in Eq.\ (\ref{exp}) $
x$ and $x_e $ differ maximum on a quantity $ O(\omega) $. This can be
seen from Eq.\ (\ref{the-eq4}) which contains the exact rate $ \lambda $
so it generates the accepted arrival $ x=x_{e_2} $ which terminates the
simulation of $e_0\rightleftharpoons e_1  $ fluctuations. State $ e_2 $
with two reversed spins should be further simulated with TBKLA. Thus,
arrival $ x $ simulated with the use of a majorizing rate is always
smaller or equal to $ x_{e_2} $ and so also is at an $ O(\omega) $
distance from $x_e $. This is illustrated in Fig.\ \ref{fig8} for
simulation with the use of a piecewise constant majorizing rate.
Further, because in Eq.\ (\ref{mu}) $ \mu(x)=O(\omega^0) $ and $ \Delta
x = x-x_{e_0}=O(\omega) $, the integral in Eq.\ (\ref{exp}) can be
approximated by a single step of the trapezoidal rule. Corrections to
the latter are of order $  (\Delta x)^2=O(\omega^2)$ which divided by $
\omega $ makes errors in Eq.\ (\ref{exp})  to be $ O(\omega) $ which is
sufficient for our level of approximation.

Thus, by neglecting $ O(\omega) $ terms the NHPP rate Eq.\
(\ref{lambdatt0}) can be found without   resource-intensive numerical
integrations and/or interpolations.

Similar to the stationary case of Sec.\ \ref{lifetime}, majorizing rate
for $ \lambda $ Eq.\ (\ref{lambdatt0}) can be obtained by neglecting
the negative second term in brackets in Eq.\ (\ref{p_1xx0}) as
\begin{equation}\label{key}
	\bar{\lambda}^0(x)=\hat{p}_1(x)r_{1\to2}(x).
\end{equation} 
However, unlike in the stationary case, the majorizing function is not
a constant but depends on $ x $ in a nontrivial manner. This would make
the thinning with its use almost as difficult as with the exact rate.

Therefore, the majorizing function $\bar{\lambda}^0 $ has been further
approximated by a piecewise constant function $ \bar{\lambda}^{st} $ as
follows \cite{nhpp-wiley,piecewise_nhpp}.  First, to implement efficient
search using the binary search algorithm, the dimensionless period $
(0,2\pi) $ has been divided into $ 2^n $ equal intervals of length $
\Delta x=2\pi/2^n $. At each point $ x_i^0=\Delta x k, k=0,...,2^n-1 $
values $ \bar{\lambda}^0_k=\bar{\lambda}^0(x_k^0) $ have been calculated
and used as majorizing constants within the intervals; to this end among
two values of $ \bar{\lambda}^0_k$ at the interval ends the larger one
has to be chosen and $ \bar{\lambda}^{st}(x) $ set equal to this value
within the interval. Next the piecewise majorizing function integrated
over $ x $ has been calculated resulting in a piecewise linear cumulative
majorizing function
\begin{equation}\label{Lambda}    
	\bar{\Lambda}(x)=\int_{0}^{x}\bar{\lambda}^{st}(x^\prime)dx^\prime
\end{equation}
which is easily calculated with the knowledge of $ \Delta x $ and of the sequence $ \{\bar{\lambda}^{st}_k\} $. It has been used to generate arrival times via an algorithm described in Refs.\ \cite{nhpp-wiley,piecewise_nhpp}. The arrivals to be thinned are generated according to Eq.\ (\ref{the-eq}) written as
\begin{equation}\label{the-eq1}
	\bar{\Lambda}(x)-\bar{\Lambda}(x_0)=\int_{x_0}^{x}\bar{\lambda}^{st}(x^\prime)dx^\prime = -\omega \ln u
\end{equation}
cast in the form
\begin{equation}\label{the-eq2}
	\bar{\Lambda}(x)=\bar{\Lambda}(x_0)-\omega \ln u.
\end{equation}
Now, given $ x_0 $, the interval $ k $ to which it belongs is found simply
as the integer part of $ x_0/\Delta x $. Then $ \bar{\Lambda}(x_0) $ is
found by simple linear interpolation within the interval. Next, having
generated $ u $ to obtain $ \bar{\Lambda}(x) $ one finds $ x $ by first
determining the interval to which $ \bar{\Lambda}(x) $ value belongs
by bracketing the it via a binary search algorithm and then finding $
x $ using inverse linear interpolation within the interval. Two steps of
the algorithm of Ref.\ \cite{piecewise_nhpp} adapted to the present case
are shown in Fig.\ \ref{fig8}.  
\section{\label{hysteresis}Hysteresis simulation}
\begin{figure}
	\includegraphics[bb=148 467 492 667,scale = 0.7]{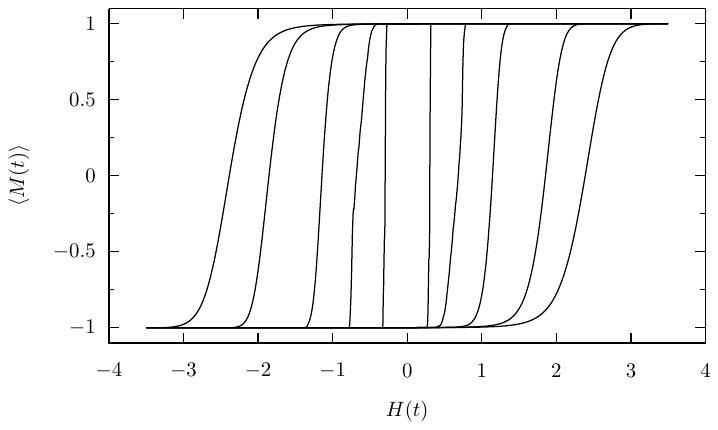}
	\caption{\label{fig9} Hysteresis loops for 2D KIM with $L=64 $
	\cite{2DIsing1998}, $ H_0 =3.5J $, and $ T = J $ for frequencies
	(in order of decreasing area)
	$ \nu/\nu_0=10^{-2}\,(10^6)$,  $10^{-2.5}\,(10^6)$,  $
	10^{-3.5}\,(10^5)$,  $ 10^{-5}\,(10^4)$, and $ 10^{-10}
	\,(6)$;  in parentheses is shown the number of simulated
	loops.}
\end{figure}

Because at sufficiently low frequencies the hysteresis loops shrink,
the hysteretic behavior eventually becomes stochastic. Its exhaustive
description requires many statistics to be gathered and analyzed which
necessitates acquisition of large volumes of data \cite{2DIsing1998}. As
it poses nontrivial storage problems while our primary goal is the
algorithms performance, in our simulations we gathered only a simple
statistic of average hysteresis loops. Several of them are shown in Fig.\
\ref{fig9}. The simulation time of one loop scales inversely proportional
to the frequency, therefore averages have been performed over $ 10^6 $
loops at the highest frequencies to only 6-10 loops at the lower end of
the frequency range due to computer time restrictions. Data have been
gathered in $10^{12} $ bins of equal width.

The loops presented in Fig.\ \ref{fig9} were simulated both with the use
of TNA and with pure TBKLA (except at the lowest frequency where TBKLA
was too slow) and no difference was discernible at the scale of the
drawing. This was a test of our additional approximation in using TNA
where for simplicity we set  magnetizations to their saturated values
during $ e_0\leftrightharpoons e_1 $ fluctuations. At equilibrium the
contributions due to one flipping spin can be roughly estimated as $
e^{-2qJ\beta}=O(10^{-4})$ at $ T=J $. However, in the dynamic case
it would depend on the relative time that the algorithm spends in
the fluctuation regime because in pure TBKLA used between the spin
fluctuations all contributions have been taken into account. At $ T=0.5J
$ the neglected contribution would be $ O(10^{-7}) $ which we consider
irrelevant in our simulations. However, if necessary, they can be taken
into account exactly, though at a cost of additional computations.

\begin{figure}
	\includegraphics[bb=149 463 484 669,scale=0.7]{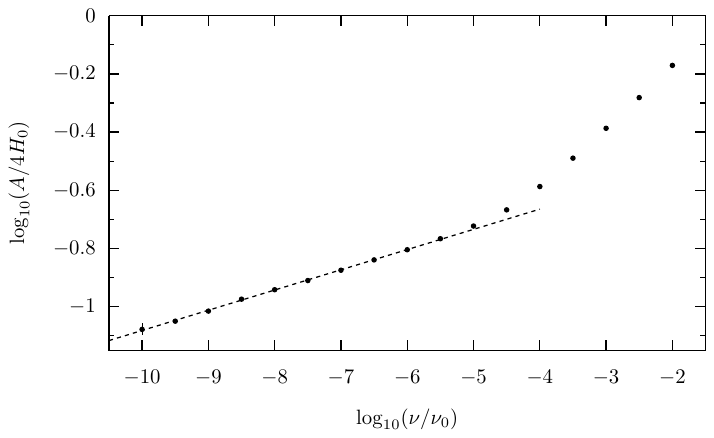}
	\caption{\label{fig10} Black circles---simulated frequency
	dependence of the  hysteresis loop area $ A $ for a KIM with
	parameters as in Fig.\ \ref{fig9}; dashed line---power-law fit
	to dependence $ \propto\omega^\gamma $ with $ \gamma\simeq0.07
	$ at low frequencies. The error bar at the leftmost point is
	the standard deviation $ \sim5\% $ which for 6 simulated loops
	amounts to standard error $ \sim2\% $ which we assess to be the
	largest error among all data points.}
\end{figure}

The frequency dependence of the loop areas for $ T=J $ case is shown in
Fig.\ \ref{fig10}. The area dispersion has been calculated only for
the lowest frequency where all six loops have been measured. Arguably,
the resulting error is the largest one because at higher frequencies the
number of simulated loops grew up to $10^6$ at the higher frequencies. The
change of the curve slope seen in Fig.\ \ref{fig10} has been attributed
to transition from stochastic to the deterministic regime as the dynamic
coercive field grows \cite{2DIsing1998,rikvold1994}.

As can be seen in Fig.\ \ref{fig3}, the TNA performance at $ T=J $ is
only about one order of magnitude faster than that of simple TBKLA with
constant majorization. We see two plausible causes for that. First, the
closeness of $ H_0 =3.5J$ to $ H_{DSP} (0)=4J$ makes the purely constant
majorization inefficient so the piecewise constant majorization should
accelerate the TBKLA part of the algorithm. Second, at higher temperatures
the number of flipped spins in the ground state grows. This can enhance
fluctuations within a set of states with more reversed spins than one
so the use of $ s>2 $ TNA may significantly improve the algorithm
performance \cite{novotny1995}.

\begin{figure}
	\includegraphics[bb=149 463 484 669,scale=0.7]{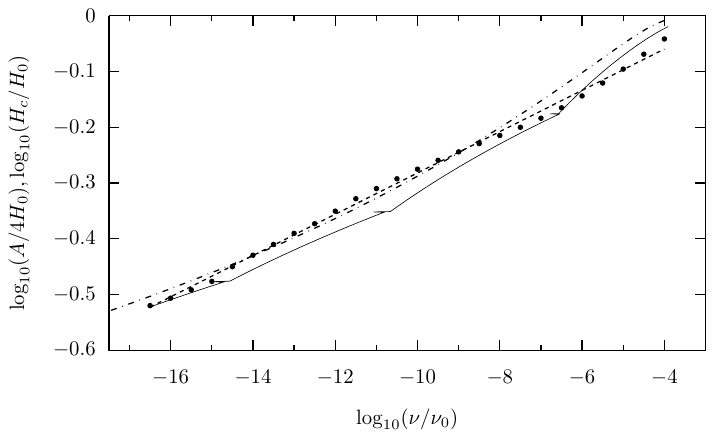}
	\caption{\label{fig11}Dots---simulated frequency
	dependence of the  hysteresis loop area of 2D KIM  at  $
	T=0.5J $ with $ H_0=1.5J $; dashed line---power-law fit to $
	\sim\omega^\gamma $ with $ \gamma\simeq0.037 $; dashed-dotted
	line---frequency dependence on the dynamic coercive
	field $ H_c $ Eq.\ (\ref{nu-v-H}); solid line---adiabatic
	approximation \cite{2DIsing1998} with the use of Eqs.\
	(\ref{rate-t})---(\ref{gamma}).}
\end{figure}

As expected, at $ T=0.5J $ it has been possible to simulate much lower
frequencies because acceleration provided by TNA is considerably higher,
as can be seen from comparison of lower and upper curves in Figs.\
\ref{fig3} and \ref{fig4}. The simulation data shown in
Fig.\ \ref{fig11} have been obtained by averaging loop areas over
$10-10^6 $ hysteresis cycles with the number growing from the lowest to
the highest frequencies. As in the case $ T=J $, at the lowest frequencies
the statistics have been gathered over individual loops and the standard
deviation has been found to be $\sim1.5\%$, hence, the standard error less
than the size of symbols in Fig.\ \ref{fig11}. As  will be shown
in the next subsection, the dispersion weakly grows with the growing
frequency.  But because the number of simulated loops has been growing
faster the estimated errors at higher frequencies has been assessed to
be negligible.

The lowest frequency $ \nu/\nu_0=10^{-16.5}=3.16\cdot10^{-17} $
simulated at $ T=0.5J $  would correspond  in physical terms to $
\nu\sim 30 $~nanohertz if the lowest estimated $ \nu_0 =1$~GHz  would
be used which is much smaller than in previous hysteresis simulations
\cite{2DIsing1998}. Going to even lower frequencies meets not only
with the problem of the fast growing computation time ($\sim16$~h per
period) but also with problems of arithmetic precision. As has been
argued earlier, from Eq.\ (\ref{the-eq4}) it follows that a typical step
size $\Delta x = O(\omega) $. Thus, to pass an oscillation period one
would have to add $ \Delta x $ to $ x $ of order of the period length
$ 2\pi $. But $ \omega/2\pi=\nu $ so with $ \nu< 3.16\cdot10^{-17}$
we are already below the double precision machine epsilon value $
2.22\cdot10^{-16} $. Thus, to simulate smaller frequencies a higher
precision arithmetic may be needed.
\subsection{\label{lowT}Low temperature theory}
Because $ T=J $ is a relatively high temperature, theoretical
interpretation of the data in Fig.\ \ref{fig10} should
be sought within the approach of Ref.\ \cite{2DIsing1998}
which, however, requires some heuristic input. On the
other hand, in Refs.\ \cite{novotny2002-3D,nita2003} it was
found that fully analytic low temperature theories of Refs.\
\cite{SD-rates,Shneidman_2003,nita2003,Shneidman_2005} can be valid up
to $ T=0.5-0.6J $.  Though the nucleation theories in these studies,
in particular, Eq.\ (\ref{tau}), were developed for the stationary case,
they can be adapted to a low frequency hysteresis within the adiabatic
approximation  \cite{2DIsing1998}. It should be noted that in the above
theories it was assumed that the system is in a metastable state while
during hysteresis it spends most of the time in stable states. Therefore,
the range of validity of Eq.\ (\ref{tau}) in our case is restricted to
such values of $ \tau $ when the external field is directed oppositely
to the system magnetization. Besides, because at low frequencies the
magnetization switch predominantly occurs within each half-period
\cite{2dIsing}, the external field Eq.\ (\ref{Ht}) will be directed
oppositely to it when $ \omega t=2\pi n +\pi/2+\omega\tau $ where $ n $
is integer and $ 0<\omega\tau <\pi $ where for concreteness we assumed
that initially the system magnetization was negative. Within this time
interval according to Eq.\ (\ref{Ht})
\begin{equation}\label{h0sin}
	H(\tau)=H_0\sin (\omega\tau)>0.
\end{equation}

In 2D case at a frequency sufficiently low for the adiabatic approximation
to be adequate \cite{droplet-theory98,2DIsing1998} the rate  of a
droplet nucleation $ \lambda_n $ can be estimated as the inverse of
the average nucleation time Eq.\ (\ref{tau}) which for 2D KIM reads
\cite{SD-rates,nita02,nita2003,novotny2002-3D}
\begin{equation}\label{rate-t}
	\lambda_n(\tau)=(L^2/A_2)e^{-\Gamma_2(\tau)/T}
\end{equation}
where \cite{Shneidman_2005}
\begin{eqnarray}
	\label{gamma-a}
	A_2 &=& 3/(8l_2 - 4),\\
	\Gamma_2(\tau) &=& 8Jl_2 - B_2H(\tau),\\
	\label{gamma}
	B_2 &=&2(l_2^2 - l_2 + 1),
\end{eqnarray}
and
\begin{equation}\label{l}
	l_2(\tau) = \left[2J/H(\tau)\right]+1. 
\end{equation}	
Here the square brackets denote the integer part which means that $
l_2(\tau) $ is discontinuous at integer values of $ 2J/H(\tau) $ while
being $ \tau $-independent in-between.

\begin{figure}
	\includegraphics[bb=148 463 486 669,scale = 0.7]{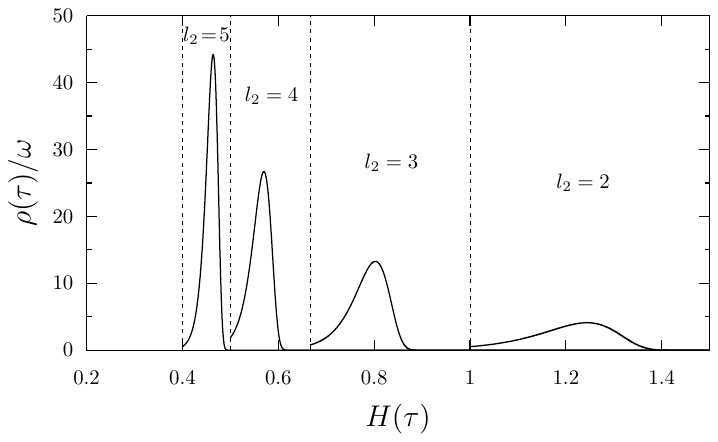}
	\caption{\label{fig12}Probability density rate of the
	magnetization switch Eq.\ (\ref{1st-arrival}) with $
	\lambda_n(\tau) $ Eq.\ (\ref{rate-t}). Only the first
	quarter-period after the field reversal at $ \tau=0 $ is
	shown. The curves from right to left are drawn, respectively, for
	$ l_2=2,3,4,5 $ where the corresponding distribution maxima occur
	at frequencies $ \nu/\nu_0=10^{-5}, 10^{-9},10^{-13},10^{-16}$
	in the same order. The values correspond to four segments of
	the solid line in Fig.\ \ref{fig11}.}
\end{figure}

Eqs.\ (\ref{tau}) and (\ref{rate-t}) have been derived for decay
of metastable states that appear after an instant reversal of the
external field to favor the spin orientation opposite to the current
magnetization. Therefore, in the case of hysteresis it is expected to
be valid only  during the first half-period after the field reversal,
say, at $ \tau=0 $, because the half-periods in the stable states are
beyond the region of validity of the equations. That is, the  underlying
picture should be that of the magnetization switching occurring every
half-period which can be observed in hysteresis with sufficiently large
periods (see Ref.\ \cite{2DIsing1998} and below).  Probability density can
be described by Eq.\ (\ref{1st-arrival}) with $ t_0=0 $ and $ t=\tau $ as
\begin{equation}\label{tau-s}
\rho(\tau)=\lambda_n(\tau)e^{-\int_0^\tau\lambda_n(\tau^\prime)d\tau^\prime}.
\end{equation}
In Fig.\ \ref{fig12} $ \rho(\tau) $ is plotted within the first
quarter-period of the oscillation where the one-to-one correspondence
between the sine function and its argument has made possible to draw
the $ \rho-H $ dependence. The integrated probability within each of
the curves shown in Fig.\ \ref{fig12} has been found to be different
from unity on less than one on $\lesssim 10^{-4} $ which supports the
qualitative picture of magnetization switching every half-period. As
can be seen, the probability rate distributions at small frequencies are
quite narrow so the average value of the dynamic coercive field $ H_c $
can be estimated by the Laplace integration \cite{bender1999advanced}
as $ H_c\equiv H(\tau_c)=H_0\sin\omega\tau_c $ where $ \tau_c $ is
the most probable switching time at the maximum of $ \rho(\tau) $ Eq.\
(\ref{tau-s}). From $ \rho^\prime(\tau_c) =0$ (prime means derivative
over $ \tau_c $) one finds
\begin{equation}\label{lambda-prime}
	\lambda^\prime_n(\tau_c)=\lambda_n^2(\tau_c)
\end{equation}
or 
\begin{equation}\label{lambda-prime2}
	[\lambda_n^{-1}(\tau_c)]^\prime=-1.
\end{equation}
Substituting this in Eqs.\ (\ref{rate-t}) and (\ref{lambda-prime2})
one finds an explicit equation for $ \tau_c $
\begin{equation}\label{the-eq0}
	(A_2/L^2)B_2 (H_0/T)\cos(\omega \tau_c)e^{\Gamma(\tau_c)/T}=1/\omega
\end{equation}
valid within the time intervals defined by Eq.\ (\ref{l}) where $ l_2,A_2,
$ and $ B_2 $ remain constant. Eq.\ (\ref{the-eq0}) can be cast in the
form   convenient for drawing plots shown in Fig.\ \ref{fig11}
\begin{equation}\label{nu-v-H}
	\nu = \frac{L^2T\exp[(B_2H_c-8Jl_2)/T]}{2\pi A_2B_2H_0\sqrt{1-(H_c/H_0)^2}}
\end{equation}
where the solid line is drawn according to this equation with the use
of Eqs.\ (\ref{gamma-a})-(\ref{gamma}). The cusps at integer values of $
l_2 $ has been drawn to show the infinitely narrow spikes in the values
of $ A_0 $ at zero temperature; at finite temperatures they will broaden
\cite{nita2003}. So with the broadening taken into account the theoretical
curves would approach the simulated data even closer. Furthermore,
the use of an average over $ \tau $ with the probability density $\rho
$ instead of using only the maximum value should smear the remaining
cusps in the curves. This suggests that low-frequency hysteresis at low
temperatures may be amenable to analytic treatment along the lines of
Refs.\ \cite{2DIsing1998,nita2003}.

As $ \omega\to0 $ $ H_c $ also tends to zero so the step separation in
Eq.\ (\ref{l}) becomes small and can be accurately approximated by a
continuous function
\begin{equation}\label{l-approx}
	l_2\simeq 2J/H_c+0.5.
\end{equation}
Substituting this in Eqs.\ (\ref{gamma-a})-(\ref{gamma}) and
(\ref{nu-v-H}) one obtains an analytic expression plotted as the
dash-dotted curve in Fig.\ \ref{fig11}. It can be easily analyzed
at small $ H_c $ and the asymptotic behavior $ A\propto 1/\ln\omega $
\cite{droplet-theory98} can be found. Thus, despite the good agreement
of the simulated data with the power-law function seen in Figs.\
\ref{fig10} and \ref{fig11}, our calculation confirm the conclusion
of Ref.\ \cite{2DIsing1998} that this is just a transient phenomenon
and the true asymptotic is the inverse  logarithm of the frequency.
\section{Summary}
In this paper we have suggested a general thinning algorithms
(GTA) fore Monte Carlo simulations of hysteresis as well
as of the decay of metastable states in the kinetic Ising
models with the Glauber dynamics. This has been achieved by
first extending the  Bortz-Kalos-Lebowitz algorithm (BKLA)
\cite{n-fold} and the Novotny algorithm (NA) \cite{novotny1995}
to the case of time-dependent Hamiltonians and by accelerating
the generation of arrival times in the ensuing non-homogeneous
Poisson processes with the use of the thinning method
\cite{thinning,streit2010poisson,nhpp-wiley,ross2010introduction,piecewise_nhpp}.

Among three implementations of GTA that we have developed the most
useful and easiest to implement is the thinned BKLA (TBKLA) which can
be efficient in systems of any size and is reasonably fast at high to
moderately low frequencies of practical interest \cite{hyperthermia} and
at temperatures where the Metropolis algorithm becomes ineffective. We
have illustrated TBKLA using the simplest  majorization by a constant
rate which is efficient at low amplitudes of the magnetic field. But
piecewise constant majorization suitable for arbitrary external field
strength would be similarly universal and easily implementable.

The efficiency of thinned NA (TNA) has been demonstrated by
simulating the decay of metastable states in the stationary case
\cite{rikvold1994,novotny2002-3D}.  With the use of $ s=2 $ TNA, where $
s-1 $ is the number of transient states used in TNA \cite{novotny1995}
we have succeeded in simulating up to 177 orders of magnitude larger
lifetimes compared to those simulated in Ref.\ \cite{novotny2002-3D}
within $ s\geq3 $ NA. We have extended TNA for oscillating external field
via thinning by a piecewise-constant majorization function. The Floquet's
theory assures that all computations can be reduced to one oscillation
period thus reducing the storage requirements. This has been implemented
via approximations that we have used to simplify computations. We assess
the approximations to be accurate for all practical purposes but if
necessary they can be improved to any desired precision, though with a
computational cost that may degrade the efficiency.

Our simulations have been restricted to relatively small nanostructures
where the single-domain (SD) nucleation dominates. In principle,
similar to Ref.\ \cite{PRE08} we could have separated the system into
non-overlapping boxes small enough to have the single-spin fluctuations
more probable thus making operative the physics underlying TNA. In
practice, however, this would pose severe combinatorial problems of
tracking the appearance of not only spin pairs within some of the boxes
but also the spin flips on the external boundaries that would cause
modification of flipping rates within the boxes. Still, restriction of
the problem to only two boxes may be feasible. This can be of interest
for simulation of nanomagnets for hyperthermia applications where the
optimum performance was found to occur at the point of transition from
SD to many-domain nucleation regime \cite{hyperthermia}. It should be
pointed out that NA and TNA remain exact in systems of any sizes, only
the probability of occurrence of the processes they describe would be
reduced in large systems. Therefore, an alternative to dealing with the
boxes may be the extension of TNA to $ s >2 $ which should make possible
simulation of larger systems and/or at higher temperatures.

Because of additional computations needed in the implementation
of thinning algorithms, their efficiency strongly depends on
temperature. We have shown that the highest acceleration is achieved
at the lowest temperatures. However, our  simulations of hysteresis
have been carried out at relatively high temperatures corresponding
roughly to region between 20\% and 40\% of the critical temperature
of the 2D Ising model.  This choice was deliberate because in Refs.\
\cite{SD-rates,Shneidman_2003,nita2003,Shneidman_2005,prefactors2003,%
novotny2002-3D}
and in the present study it has been found that below about 25\% of
the critical temperature analytic theories show good agreement with the
simulated data so in principle  at low temperatures reliable analytic
theories of hysteresis should be feasible. At higher temperatures,
however, both in the present study and In Ref.\ \cite{2DIsing1998}
the agreement was much poorer so we expect that thinning algorithms
would be the most useful for temperatures above $ \sim 0.25$ of the
critical temperature.  For some commonly used ferromagnetic materials
this range includes the room temperature and the temperatures of interest
for hyperthermia applications \cite{hyperthermia} so it is hoped that
suitably modified thinning algorithms  can be useful in simulations of
magnetic nanostructures of practical interest.
\begin{acknowledgments}
We express our gratitude to B\'eatrice Masson for help with acquisition
of some bibliographical sources. One of us (V.I.T.) is thankful to the
Dreyss\'e family for their hospitality.
\end{acknowledgments}
\appendix
\section{\label{appa}The first arrival time}
The probability distribution of the first arrival after $ t_0 $
can be derived as follows. Let the probability that no arrivals
occurred in the interval $ (t_0,t) $ be $ \bar{p}(t,t_0) $. Then the
probability of no arrivals at time $ t+\Delta t $ is $\bar{p}(t+\Delta
t,t_0)=\bar{p}(t,t_0)[1-\lambda(t)\Delta t]$ where the second term in the
brackets accounts for the probability of arrival within time interval $
(t,t+\Delta t) $. With the initial condition $ \bar{p}(t_0,t_0)=1$
the solution of the ensuing differential equation is
\begin{equation}\label{bar-p}
	\bar{p}(t,t_0) = e^{-\int_{t_0}^{t}\lambda(t^\prime)dt^\prime}.
\end{equation} 
Hence, the probability that the arrival did occur within time interval $ (t_0,t) $ is
\begin{eqnarray}\label{cpd}
	f(t,t_0) = 1-\bar{p}(t,t_0)&=&1-e^{-\int_{t_0}^{t}\lambda(t^\prime)dt^\prime}\nonumber\\
	&=&\int_{t_0}^{t}\rho(t^\prime,t_0)dt^\prime
\end{eqnarray}
where the  equality on the second line reflects the fact that in $ f $
all occurrences of the event anywhere within the time interval have
been taken into account. So $ f $ is the cumulative distribution for
some probability density function $ \rho $.

The task of simulating the arrival event is solved by the theorem stating
that the values of stochastic  function $ f $ in Eq.\ (\ref{cpd}) are
homogeneously distributed within the interval $ (0,1) $ (see, e.g., Ref.\
\cite{cpd-theorem}, Theorem 2.1.10). So one has to generate the uniform
variate $ u\in (0,1)$ and determine arrival time $ t $ by solving equation
\begin{equation}\label{f=u}
	f(t,t_0)=u.
\end{equation}
Here we note that in PP the arrivals cannot be simultaneous, so there
always is the first arrival which terminates the no-arrival time
interval. Therefore, by differentiating Eq.\ (\ref{cpd}) with respect
to $ t $ one finds that $ \rho $ is the rate of the first arrival
\cite{streit2010poisson}
\begin{equation}\label{1st-arrival}
	\rho(t,t_0)=\lambda(t)e^{-\int_{t_0}^{t}\lambda(t^\prime)dt^\prime}
\end{equation}
which is further confirmed by its unit normalization. This expression
generalizes to NHPP the exponential distribution of interarrival intervals
of the conventional PP with constant $ \lambda $. Moreover, NHPP can be
mapped onto PP by a nonlinear transformation \cite{streit2010poisson}.
\section{\label{transition}The transition matrix}
The rate matrix in Eq.\ (\ref{R}) has the form
\begin{equation}\label{Ra}
	R(t)=\begin{bmatrix}
		-a_0(t)&a_1(t)\\ a_0(t)&-a_1(t)
	\end{bmatrix},
\end{equation} 
where explicit expressions for functions $ a_i $ can be found in Eq.\ (\ref{R}). 
The $ 2\times2 $ transition matrix is defined as
\begin{equation}
	P(t,t_0)=\begin{bmatrix}
		1- p_0(t,t_0)& p_1(t,t_0)\\  p_0(t,t_0)&1- p_1(t,t_0)
	\end{bmatrix}=\left[\psi_1 \psi_0\right].
	\label{Ptt0}
\end{equation} 
Its columns are vectors $ \psi_1 $ and $ \psi_0 $ so according to Eq.\
(\ref{dot_psi}) it satisfies equation
\begin{equation}\label{ddtP}
	\frac{d}{dt}P(t,t_0)=R(t)P(t,t_0)
\end{equation}
with the initial condition
\begin{equation}\label{P0}
	P(t_0,t_0)=I.
\end{equation}
With the use of Eq.\ (\ref{ddtP}) it can be shown that matrix $ P $
also satisfies functional equation \cite{brockett1970}
\begin{equation}
	{ P}(t,t_0)={ P}(t,t_1){ P}(t_1,t_0)
	\label{group}
\end{equation}
where $ t\geq t_1\geq t_0 $.  Now assuming that the range of the time
variable is $ [0,\infty)$, it is easy to see from Eq.\ (\ref{group})
that $ { P}(t,t_0) $ can be expressed in terms of $ \hat{P}(t)\equiv {
P}(t,0) $ as
\begin{equation}\label{P/P}
	{ P}(t,t_0)=	\hat{P}(t) \hat{P}^{-1}(t_0).
\end{equation}
Thus, a two-parameter matrix $ P $ can be calculated with the use of a
one-parameter matrix $ \hat{P} $ which can reduce computational overhead
in KMC simulations when an analytical solution is not available. In fact,
from Eq.\ (\ref{Ptt0}) it can be seen that only one matrix element of $
P $ is needed in Eq.\ (\ref{lambdatt0}).
\begin{equation}
	\dot{\hat{ p}}_k=-(a_0+a_1)\hat{ p}_k+a_k
	\label{dot_phi12}
\end{equation}
where $ \hat{ p}_k(t)=p_k(t,0) $ (the arguments $ t $ of all terms
in the equation have been omitted for brevity). The solution of Eq.\
(\ref{dot_phi12}) reads
\begin{equation}\label{p_it}
	\hat{p}_k(t)=\int_{0}^{t}\exp\left[-\int_{t^\prime}^{t}\mu(t_1)
	dt_1\right]a_k(t^\prime)dt^\prime,
\end{equation} 
where
\begin{equation}\label{mu}
	\mu(t)=a_0(t)+a_1(t).
\end{equation}
To invert $ \hat{P}(t_0) $ in Eq.\ (\ref{P/P}) we will need its determinant 
\begin{equation}\label{det}
	\det \hat{P}(t) =1-\hat{p}_0(t)-\hat{p}_1(t)= \exp\left[-\int_{0}^{t}\mu(t^\prime)dt^\prime\right]
\end{equation}
where the second equality (a particular case of the Abel-Jacobi-Liouville
theorem \cite{brockett1970}) has been obtained with the use of Eqs.\
(\ref{Ptt0}) and (\ref{p_it}). Now by using Eq.\ (\ref{P/P}) one gets
the probability
\begin{equation}\label{p_1tt0}
	p_1(t,t_0)=\hat{p}_1(t)-\exp\left[-\int_{t_0}^{t}\mu(t^\prime)dt^\prime\right]\hat{p}_1(t_0)
\end{equation}
that enters NHPP rate Eq.\ (\ref{lambdatt0}). An important observation
concerning this expression is that dependence on $ t_0 $ on the rhs is
confined to the second term which is not positive because $ \hat{p}_1\geq0
$. Thus, the first term  which depends only on $ t $ is a majorizing
function to $ \lambda(t) $ Eq.\ (\ref{lambdatt0}). This means that a
computationally simple one-parameter majorization is possible in this
problem.
\section{\label{Laplace}The Laplace integral}
Eq.\ (\ref{change}) is a Laplace integral of the form 
\begin{eqnarray}\label{laplace}
	I(K)&=&\int_{a}^{b}f(x)e^{K\phi(x)}dx\nonumber\\& =& \frac{1}{K}\int_{a}^{b}\frac{f(x)}{\phi\prime(x)}d\left[e^{K\phi(x)}\right]
\end{eqnarray}
with $ \phi(x) $ reaching its maximum value---zero in our case---at
the upper integration boundary.  Integrating by parts one finds (Ref.\
\cite{bender1999advanced}, Ch.\ 6.3)
\begin{equation}\label{asympt}
I(K)\simeq \frac{1}{K}\frac{f(b)}{\phi\prime(b)}e^{K\phi(b)}+O\left(\frac{1}{K^2}\right).	
\end{equation}
Further terms of the asymptotic expansion can also be calculated \cite{bender1999advanced}.

\begin{thebibliography}{47}%
\makeatletter
\providecommand \@ifxundefined [1]{%
 \@ifx{#1\undefined}
}%
\providecommand \@ifnum [1]{%
 \ifnum #1\expandafter \@firstoftwo
 \else \expandafter \@secondoftwo
 \fi
}%
\providecommand \@ifx [1]{%
 \ifx #1\expandafter \@firstoftwo
 \else \expandafter \@secondoftwo
 \fi
}%
\providecommand \natexlab [1]{#1}%
\providecommand \enquote  [1]{``#1''}%
\providecommand \bibnamefont  [1]{#1}%
\providecommand \bibfnamefont [1]{#1}%
\providecommand \citenamefont [1]{#1}%
\providecommand \href@noop [0]{\@secondoftwo}%
\providecommand \href [0]{\begingroup \@sanitize@url \@href}%
\providecommand \@href[1]{\@@startlink{#1}\@@href}%
\providecommand \@@href[1]{\endgroup#1\@@endlink}%
\providecommand \@sanitize@url [0]{\catcode `\\12\catcode `\$12\catcode
  `\&12\catcode `\#12\catcode `\^12\catcode `\_12\catcode `\%12\relax}%
\providecommand \@@startlink[1]{}%
\providecommand \@@endlink[0]{}%
\providecommand \url  [0]{\begingroup\@sanitize@url \@url }%
\providecommand \@url [1]{\endgroup\@href {#1}{\urlprefix }}%
\providecommand \urlprefix  [0]{URL }%
\providecommand \Eprint [0]{\href }%
\providecommand \doibase [0]{https://doi.org/}%
\providecommand \selectlanguage [0]{\@gobble}%
\providecommand \bibinfo  [0]{\@secondoftwo}%
\providecommand \bibfield  [0]{\@secondoftwo}%
\providecommand \translation [1]{[#1]}%
\providecommand \BibitemOpen [0]{}%
\providecommand \bibitemStop [0]{}%
\providecommand \bibitemNoStop [0]{.\EOS\space}%
\providecommand \EOS [0]{\spacefactor3000\relax}%
\providecommand \BibitemShut  [1]{\csname bibitem#1\endcsname}%
\let\auto@bib@innerbib\@empty
%</preamble>
\bibitem [{\citenamefont {Macy}\ \emph {et~al.}(2024)\citenamefont {Macy},
  \citenamefont {Szymanski},\ and\ \citenamefont
  {Ho\l{}yst}}]{interdisciplinary24}%
  \BibitemOpen
  \bibfield  {author} {\bibinfo {author} {\bibfnamefont {M.~W.}\ \bibnamefont
  {Macy}}, \bibinfo {author} {\bibfnamefont {B.~K.}\ \bibnamefont
  {Szymanski}},\ and\ \bibinfo {author} {\bibfnamefont {J.~A.}\ \bibnamefont
  {Ho\l{}yst}},\ }\bibfield  {title} {\bibinfo {title} {The ising model
  celebrates a century of interdisciplinary contributions},\ }\href
  {https://doi.org/10.1038/s44260-024-00012-0} {\bibfield  {journal} {\bibinfo
  {journal} {npj Complexity}\ }\textbf {\bibinfo {volume} {1}},\ \bibinfo
  {pages} {10} (\bibinfo {year} {2024})}\BibitemShut {NoStop}%
\bibitem [{\citenamefont {Ducastelle}(1991)}]{ducastelle}%
  \BibitemOpen
  \bibfield  {author} {\bibinfo {author} {\bibfnamefont {F.}~\bibnamefont
  {Ducastelle}},\ }\href@noop {} {\emph {\bibinfo {title} {Order and Phase
  Stability in Alloys}}}\ (\bibinfo  {publisher} {North-Holland},\ \bibinfo
  {address} {Amsterdam},\ \bibinfo {year} {1991})\BibitemShut {NoStop}%
\bibitem [{\citenamefont {Bruscolini}\ \emph {et~al.}(2007)\citenamefont
  {Bruscolini}, \citenamefont {Pelizzola},\ and\ \citenamefont
  {Zamparo}}]{BPZ}%
  \BibitemOpen
  \bibfield  {author} {\bibinfo {author} {\bibfnamefont {P.}~\bibnamefont
  {Bruscolini}}, \bibinfo {author} {\bibfnamefont {A.}~\bibnamefont
  {Pelizzola}},\ and\ \bibinfo {author} {\bibfnamefont {M.}~\bibnamefont
  {Zamparo}},\ }\bibfield  {title} {\bibinfo {title} {Rate determining factors
  in protein model structures},\ }\href@noop {} {\bibfield  {journal} {\bibinfo
   {journal} {Phys. Rev. Lett.}\ }\textbf {\bibinfo {volume} {99}},\ \bibinfo
  {pages} {038103} (\bibinfo {year} {2007})}\BibitemShut {NoStop}%
\bibitem [{\citenamefont {Aguilera}\ \emph {et~al.}(2021)\citenamefont
  {Aguilera}, \citenamefont {Moosavi},\ and\ \citenamefont
  {Shimazaki}}]{aguilera_unifying_2021}%
  \BibitemOpen
  \bibfield  {author} {\bibinfo {author} {\bibfnamefont {M.}~\bibnamefont
  {Aguilera}}, \bibinfo {author} {\bibfnamefont {S.~A.}\ \bibnamefont
  {Moosavi}},\ and\ \bibinfo {author} {\bibfnamefont {H.}~\bibnamefont
  {Shimazaki}},\ }\bibfield  {title} {\bibinfo {title} {A unifying framework
  for mean-field theories of asymmetric kinetic {Ising} systems},\ }\href
  {https://doi.org/10.1038/s41467-021-20890-5} {\bibfield  {journal} {\bibinfo
  {journal} {Nat. Commun.}\ }\textbf {\bibinfo {volume} {12}},\ \bibinfo
  {pages} {1197} (\bibinfo {year} {2021})}\BibitemShut {NoStop}%
\bibitem [{\citenamefont {Kirkpatrick}\ \emph {et~al.}(1983)\citenamefont
  {Kirkpatrick}, \citenamefont {Gelatt},\ and\ \citenamefont
  {Vecchi}}]{Kirkpatrick1983OptimizationBS}%
  \BibitemOpen
  \bibfield  {author} {\bibinfo {author} {\bibfnamefont {S.}~\bibnamefont
  {Kirkpatrick}}, \bibinfo {author} {\bibfnamefont {C.~D.}\ \bibnamefont
  {Gelatt}},\ and\ \bibinfo {author} {\bibfnamefont {M.~P.}\ \bibnamefont
  {Vecchi}},\ }\bibfield  {title} {\bibinfo {title} {Optimization by simulated
  annealing},\ }\href@noop {} {\bibfield  {journal} {\bibinfo  {journal}
  {Science}\ }\textbf {\bibinfo {volume} {220}},\ \bibinfo {pages} {671}
  (\bibinfo {year} {1983})}\BibitemShut {NoStop}%
\bibitem [{\citenamefont {Landau}\ and\ \citenamefont
  {Binder}(2009)}]{landau2009guide}%
  \BibitemOpen
  \bibfield  {author} {\bibinfo {author} {\bibfnamefont {D.}~\bibnamefont
  {Landau}}\ and\ \bibinfo {author} {\bibfnamefont {K.}~\bibnamefont
  {Binder}},\ }\href@noop {} {\emph {\bibinfo {title} {A Guide to Monte Carlo
  Simulations in Statistical Physics}}}\ (\bibinfo  {publisher} {Cambridge
  University Press},\ \bibinfo {year} {2009})\BibitemShut {NoStop}%
\bibitem [{\citenamefont {Chatterjee}\ and\ \citenamefont
  {Vlachos}(2007)}]{KMCreview07}%
  \BibitemOpen
  \bibfield  {author} {\bibinfo {author} {\bibfnamefont {A.}~\bibnamefont
  {Chatterjee}}\ and\ \bibinfo {author} {\bibfnamefont {D.~G.}\ \bibnamefont
  {Vlachos}},\ }\bibfield  {title} {\bibinfo {title} {An overview of spatial
  microscopic and accelerated kinetic monte carlo methods},\ }\href@noop {}
  {\bibfield  {journal} {\bibinfo  {journal} {J. Comput.-Aided Mater. Des.}\
  }\textbf {\bibinfo {volume} {14}},\ \bibinfo {pages} {253} (\bibinfo {year}
  {2007})}\BibitemShut {NoStop}%
\bibitem [{\citenamefont {Binder}\ and\ \citenamefont
  {Virnau}(2016)}]{binder-nucleation}%
  \BibitemOpen
  \bibfield  {author} {\bibinfo {author} {\bibfnamefont {K.}~\bibnamefont
  {Binder}}\ and\ \bibinfo {author} {\bibfnamefont {P.}~\bibnamefont
  {Virnau}},\ }\bibfield  {title} {\bibinfo {title} {{Overview: Understanding
  nucleation phenomena from simulations of lattice gas models}},\ }\href
  {https://doi.org/10.1063/1.4959235} {\bibfield  {journal} {\bibinfo
  {journal} {The Journal of Chemical Physics}\ }\textbf {\bibinfo {volume}
  {145}},\ \bibinfo {pages} {211701} (\bibinfo {year} {2016})}\BibitemShut
  {NoStop}%
\bibitem [{\citenamefont {Metropolis}\ \emph {et~al.}(1953)\citenamefont
  {Metropolis}, \citenamefont {Rosenbluth}, \citenamefont {Rosenbluth},
  \citenamefont {Teller},\ and\ \citenamefont {Teller}}]{metropolis}%
  \BibitemOpen
  \bibfield  {author} {\bibinfo {author} {\bibfnamefont {N.}~\bibnamefont
  {Metropolis}}, \bibinfo {author} {\bibfnamefont {A.~W.}\ \bibnamefont
  {Rosenbluth}}, \bibinfo {author} {\bibfnamefont {M.~N.}\ \bibnamefont
  {Rosenbluth}}, \bibinfo {author} {\bibfnamefont {A.~H.}\ \bibnamefont
  {Teller}},\ and\ \bibinfo {author} {\bibfnamefont {E.}~\bibnamefont
  {Teller}},\ }\bibfield  {title} {\bibinfo {title} {Equation of state
  calculations by fast computing machines},\ }\href
  {https://doi.org/10.1063/1.1699114} {\bibfield  {journal} {\bibinfo
  {journal} {J. Chem. Phys.}\ }\textbf {\bibinfo {volume} {21}},\ \bibinfo
  {pages} {1087} (\bibinfo {year} {1953})}\BibitemShut {NoStop}%
\bibitem [{\citenamefont {Bortz}\ \emph {et~al.}(1975)\citenamefont {Bortz},
  \citenamefont {Kalos},\ and\ \citenamefont {Lebowitz}}]{n-fold}%
  \BibitemOpen
  \bibfield  {author} {\bibinfo {author} {\bibfnamefont {A.~B.}\ \bibnamefont
  {Bortz}}, \bibinfo {author} {\bibfnamefont {M.~H.}\ \bibnamefont {Kalos}},\
  and\ \bibinfo {author} {\bibfnamefont {J.~L.}\ \bibnamefont {Lebowitz}},\
  }\href@noop {} {\bibfield  {journal} {\bibinfo  {journal} {J. Comput. Phys.}\
  }\textbf {\bibinfo {volume} {17}},\ \bibinfo {pages} {10} (\bibinfo {year}
  {1975})}\BibitemShut {NoStop}%
\bibitem [{\citenamefont {Novotny}(1995)}]{novotny1995}%
  \BibitemOpen
  \bibfield  {author} {\bibinfo {author} {\bibfnamefont {M.~A.}\ \bibnamefont
  {Novotny}},\ }\bibfield  {title} {\bibinfo {title} {Monte carlo algorithms
  with absorbing markov chains: Fast local algorithms for slow dynamics},\
  }\href {https://doi.org/10.1103/PhysRevLett.74.1} {\bibfield  {journal}
  {\bibinfo  {journal} {Phys. Rev. Lett.}\ }\textbf {\bibinfo {volume} {74}},\
  \bibinfo {pages} {1} (\bibinfo {year} {1995})}\BibitemShut {NoStop}%
\bibitem [{\citenamefont {Adam}\ \emph {et~al.}(1999)\citenamefont {Adam},
  \citenamefont {Billard},\ and\ \citenamefont {Lan\ifmmode~\mbox{\c{c}}\else
  \c{c}\fi{}on}}]{KMC1999}%
  \BibitemOpen
  \bibfield  {author} {\bibinfo {author} {\bibfnamefont {E.}~\bibnamefont
  {Adam}}, \bibinfo {author} {\bibfnamefont {L.}~\bibnamefont {Billard}},\ and\
  \bibinfo {author} {\bibfnamefont {F.}~\bibnamefont
  {Lan\ifmmode~\mbox{\c{c}}\else \c{c}\fi{}on}},\ }\bibfield  {title} {\bibinfo
  {title} {Class of monte carlo algorithms for dynamic problems leads to an
  adaptive method},\ }\href {https://doi.org/10.1103/PhysRevE.59.1212}
  {\bibfield  {journal} {\bibinfo  {journal} {Phys. Rev. E}\ }\textbf {\bibinfo
  {volume} {59}},\ \bibinfo {pages} {1212} (\bibinfo {year}
  {1999})}\BibitemShut {NoStop}%
\bibitem [{\citenamefont {Trygubenko}\ and\ \citenamefont
  {Wales}(2006)}]{trygubenko}%
  \BibitemOpen
  \bibfield  {author} {\bibinfo {author} {\bibfnamefont {S.~A.}\ \bibnamefont
  {Trygubenko}}\ and\ \bibinfo {author} {\bibfnamefont {D.~J.}\ \bibnamefont
  {Wales}},\ }\bibfield  {title} {\bibinfo {title} {{Graph transformation
  method for calculating waiting times in Markov chains}},\ }\href
  {https://doi.org/10.1063/1.2198806} {\bibfield  {journal} {\bibinfo
  {journal} {J. Chem. Phys.}\ }\textbf {\bibinfo {volume} {124}},\ \bibinfo
  {pages} {234110} (\bibinfo {year} {2006})}\BibitemShut {NoStop}%
\bibitem [{\citenamefont {Opplestrup}\ \emph {et~al.}(2006)\citenamefont
  {Opplestrup}, \citenamefont {Bulatov}, \citenamefont {Gilmer}, \citenamefont
  {Kalos},\ and\ \citenamefont {Sadigh}}]{opplestrup2006}%
  \BibitemOpen
  \bibfield  {author} {\bibinfo {author} {\bibfnamefont {T.}~\bibnamefont
  {Opplestrup}}, \bibinfo {author} {\bibfnamefont {V.~V.}\ \bibnamefont
  {Bulatov}}, \bibinfo {author} {\bibfnamefont {G.~H.}\ \bibnamefont {Gilmer}},
  \bibinfo {author} {\bibfnamefont {M.~H.}\ \bibnamefont {Kalos}},\ and\
  \bibinfo {author} {\bibfnamefont {B.}~\bibnamefont {Sadigh}},\ }\bibfield
  {title} {\bibinfo {title} {First-passage monte carlo algorithm: Diffusion
  without all the hops},\ }\href
  {https://doi.org/10.1103/PhysRevLett.97.230602} {\bibfield  {journal}
  {\bibinfo  {journal} {Phys. Rev. Lett.}\ }\textbf {\bibinfo {volume} {97}},\
  \bibinfo {pages} {230602} (\bibinfo {year} {2006})}\BibitemShut {NoStop}%
\bibitem [{\citenamefont {Tokar}\ and\ \citenamefont
  {Dreyss{\'e}}(2008)}]{PRE08}%
  \BibitemOpen
  \bibfield  {author} {\bibinfo {author} {\bibfnamefont {V.~I.}\ \bibnamefont
  {Tokar}}\ and\ \bibinfo {author} {\bibfnamefont {H.}~\bibnamefont
  {Dreyss{\'e}}},\ }\bibfield  {title} {\bibinfo {title} {Accelerated kinetic
  monte carlo algorithm for diffusion-limited kinetics},\ }\href
  {https://doi.org/10.1103/PhysRevE.77.066705} {\bibfield  {journal} {\bibinfo
  {journal} {Phys. Rev. E}\ }\textbf {\bibinfo {volume} {77}},\ \bibinfo
  {pages} {066705} (\bibinfo {year} {2008})}\BibitemShut {NoStop}%
\bibitem [{\citenamefont {Chakrabarti}\ and\ \citenamefont
  {Acharyya}(1999)}]{chakrabarti_dynamic_1999}%
  \BibitemOpen
  \bibfield  {author} {\bibinfo {author} {\bibfnamefont {B.~K.}\ \bibnamefont
  {Chakrabarti}}\ and\ \bibinfo {author} {\bibfnamefont {M.}~\bibnamefont
  {Acharyya}},\ }\bibfield  {title} {\bibinfo {title} {Dynamic transitions and
  hysteresis},\ }\href@noop {} {\bibfield  {journal} {\bibinfo  {journal} {Rev.
  Mod. Phys.}\ }\textbf {\bibinfo {volume} {71}},\ \bibinfo {pages} {847}
  (\bibinfo {year} {1999})}\BibitemShut {NoStop}%
\bibitem [{\citenamefont {Novotny}\ \emph {et~al.}(2007)\citenamefont
  {Novotny}, \citenamefont {Robb}, \citenamefont {Stinnett}, \citenamefont
  {Brown},\ and\ \citenamefont {Rikvold}}]{Novotny-nanomagnets2007}%
  \BibitemOpen
  \bibfield  {author} {\bibinfo {author} {\bibfnamefont {M.~A.}\ \bibnamefont
  {Novotny}}, \bibinfo {author} {\bibfnamefont {D.~T.}\ \bibnamefont {Robb}},
  \bibinfo {author} {\bibfnamefont {S.~M.}\ \bibnamefont {Stinnett}}, \bibinfo
  {author} {\bibfnamefont {G.}~\bibnamefont {Brown}},\ and\ \bibinfo {author}
  {\bibfnamefont {P.~A.}\ \bibnamefont {Rikvold}},\ }\bibinfo {title}
  {Finite-temperature simulations for magnetic nanostructures},\ in\ \href
  {https://doi.org/10.1007/978-3-540-49336-5_7} {\emph {\bibinfo {booktitle}
  {Magnetic Nanostructures}}},\ \bibinfo {editor} {edited by\ \bibinfo {editor}
  {\bibfnamefont {B.}~\bibnamefont {Akta{\c{s}}}}, \bibinfo {editor}
  {\bibfnamefont {F.}~\bibnamefont {Mikailov}},\ and\ \bibinfo {editor}
  {\bibfnamefont {L.}~\bibnamefont {Tagirov}}}\ (\bibinfo  {publisher}
  {Springer Berlin Heidelberg},\ \bibinfo {address} {Berlin, Heidelberg},\
  \bibinfo {year} {2007})\ pp.\ \bibinfo {pages} {97--117}\BibitemShut
  {NoStop}%
\bibitem [{\citenamefont {Giustini}\ \emph {et~al.}(2010)\citenamefont
  {Giustini}, \citenamefont {Petryk}, \citenamefont {Cassim}, \citenamefont
  {Tate}, \citenamefont {Baker},\ and\ \citenamefont {Hoopes}}]{hyperthermia}%
  \BibitemOpen
  \bibfield  {author} {\bibinfo {author} {\bibfnamefont {A.~J.}\ \bibnamefont
  {Giustini}}, \bibinfo {author} {\bibfnamefont {A.~A.}\ \bibnamefont
  {Petryk}}, \bibinfo {author} {\bibfnamefont {S.~M.}\ \bibnamefont {Cassim}},
  \bibinfo {author} {\bibfnamefont {J.~A.}\ \bibnamefont {Tate}}, \bibinfo
  {author} {\bibfnamefont {I.}~\bibnamefont {Baker}},\ and\ \bibinfo {author}
  {\bibfnamefont {P.~J.}\ \bibnamefont {Hoopes}},\ }\bibfield  {title}
  {\bibinfo {title} {Magnetic nanoparticle hyperthermia in cancer treatment},\
  }\href {https://doi.org/10.1142/S1793984410000067} {\bibfield  {journal}
  {\bibinfo  {journal} {Nano LIFE}\ }\textbf {\bibinfo {volume} {01}},\
  \bibinfo {pages} {17} (\bibinfo {year} {2010})}\BibitemShut {NoStop}%
\bibitem [{\citenamefont {Sides}\ \emph
  {et~al.}(1998{\natexlab{a}})\citenamefont {Sides}, \citenamefont {Rikvold},\
  and\ \citenamefont {Novotny}}]{2DIsing1998}%
  \BibitemOpen
  \bibfield  {author} {\bibinfo {author} {\bibfnamefont {S.~W.}\ \bibnamefont
  {Sides}}, \bibinfo {author} {\bibfnamefont {P.~A.}\ \bibnamefont {Rikvold}},\
  and\ \bibinfo {author} {\bibfnamefont {M.~A.}\ \bibnamefont {Novotny}},\
  }\bibfield  {title} {\bibinfo {title} {Stochastic hysteresis and resonance in
  a kinetic {I}sing system},\ }\href@noop {} {\bibfield  {journal} {\bibinfo
  {journal} {Phys. Rev. E}\ }\textbf {\bibinfo {volume} {57}},\ \bibinfo
  {pages} {6512} (\bibinfo {year} {1998}{\natexlab{a}})}\BibitemShut {NoStop}%
\bibitem [{\citenamefont {Zhu}\ \emph {et~al.}(2004)\citenamefont {Zhu},
  \citenamefont {Dong},\ and\ \citenamefont {Liu}}]{zhu_hysteresis_2004}%
  \BibitemOpen
  \bibfield  {author} {\bibinfo {author} {\bibfnamefont {H.}~\bibnamefont
  {Zhu}}, \bibinfo {author} {\bibfnamefont {S.}~\bibnamefont {Dong}},\ and\
  \bibinfo {author} {\bibfnamefont {J.-M.}\ \bibnamefont {Liu}},\ }\bibfield
  {title} {\bibinfo {title} {Hysteresis loop area of the ising model},\
  }\href@noop {} {\bibfield  {journal} {\bibinfo  {journal} {Phys. Rev. B}\
  }\textbf {\bibinfo {volume} {70}},\ \bibinfo {pages} {132403} (\bibinfo
  {year} {2004})}\BibitemShut {NoStop}%
\bibitem [{\citenamefont {Novotny}\ \emph {et~al.}(2002)\citenamefont
  {Novotny}, \citenamefont {Brown},\ and\ \citenamefont
  {Rikvold}}]{novotny-review2002}%
  \BibitemOpen
  \bibfield  {author} {\bibinfo {author} {\bibfnamefont {M.~A.}\ \bibnamefont
  {Novotny}}, \bibinfo {author} {\bibfnamefont {G.}~\bibnamefont {Brown}},\
  and\ \bibinfo {author} {\bibfnamefont {P.~A.}\ \bibnamefont {Rikvold}},\
  }\bibfield  {title} {\bibinfo {title} {{Large-scale computer investigations
  of finite-temperature nucleation and growth phenomena in magnetization
  reversal and hysteresis (invited)}},\ }\href
  {https://doi.org/10.1063/1.1452188} {\bibfield  {journal} {\bibinfo
  {journal} {J. Appl. Phys.}\ }\textbf {\bibinfo {volume} {91}},\ \bibinfo
  {pages} {6908} (\bibinfo {year} {2002})}\BibitemShut {NoStop}%
\bibitem [{\citenamefont {Streit}(2010)}]{streit2010poisson}%
  \BibitemOpen
  \bibfield  {author} {\bibinfo {author} {\bibfnamefont {R.}~\bibnamefont
  {Streit}},\ }\href {https://books.google.fr/books?id=KAWmFYUJ5zsC} {\emph
  {\bibinfo {title} {Poisson Point Processes: Imaging, Tracking, and
  Sensing}}}\ (\bibinfo  {publisher} {Springer US},\ \bibinfo {year}
  {2010})\BibitemShut {NoStop}%
\bibitem [{\citenamefont {Pasupathy}(2011)}]{nhpp-wiley}%
  \BibitemOpen
  \bibfield  {author} {\bibinfo {author} {\bibfnamefont {R.}~\bibnamefont
  {Pasupathy}},\ }\bibinfo {title} {Generating nonhomogeneous poisson
  processes},\ in\ \href
  {https://doi.org/https://doi.org/10.1002/9780470400531.eorms0356} {\emph
  {\bibinfo {booktitle} {Wiley Encyclopedia of Operations Research and
  Management Science}}}\ (\bibinfo  {publisher} {John Wiley \& Sons, Ltd},\
  \bibinfo {year} {2011})\ pp.\ \bibinfo {pages} {1--11}\BibitemShut {NoStop}%
\bibitem [{\citenamefont {Ross}(2010)}]{ross2010introduction}%
  \BibitemOpen
  \bibfield  {author} {\bibinfo {author} {\bibfnamefont {S.}~\bibnamefont
  {Ross}},\ }\href {https://books.google.fr/books?id=HKwN0A5-5wwC} {\emph
  {\bibinfo {title} {Introduction to Probability Models 10th Edition}}}\
  (\bibinfo  {publisher} {Academic Press},\ \bibinfo {year} {2010})\BibitemShut
  {NoStop}%
\bibitem [{\citenamefont {Harrod}\ and\ \citenamefont
  {Kelton}(2006)}]{piecewise_nhpp}%
  \BibitemOpen
  \bibfield  {author} {\bibinfo {author} {\bibfnamefont {S.}~\bibnamefont
  {Harrod}}\ and\ \bibinfo {author} {\bibfnamefont {W.~D.}\ \bibnamefont
  {Kelton}},\ }\bibfield  {title} {\bibinfo {title} {Numerical methods for
  realizing nonstationary poisson processes with piecewise-constant
  instantaneous-rate functions},\ }\href
  {https://doi.org/10.1177/0037549706065514} {\bibfield  {journal} {\bibinfo
  {journal} {SIMULATION}\ }\textbf {\bibinfo {volume} {82}},\ \bibinfo {pages}
  {147} (\bibinfo {year} {2006})}\BibitemShut {NoStop}%
\bibitem [{\citenamefont {Casella}\ and\ \citenamefont
  {Berger}(2024)}]{cpd-theorem}%
  \BibitemOpen
  \bibfield  {author} {\bibinfo {author} {\bibfnamefont {G.}~\bibnamefont
  {Casella}}\ and\ \bibinfo {author} {\bibfnamefont {R.}~\bibnamefont
  {Berger}},\ }\href {https://doi.org/10.1201/97810034562850} {\emph {\bibinfo
  {title} {Statistical Inference (2nd ed.)}}}\ (\bibinfo  {publisher} {Chapman
  and Hall/CRC},\ \bibinfo {year} {2024})\BibitemShut {NoStop}%
\bibitem [{\citenamefont {Lewis}\ and\ \citenamefont
  {Shedler}(1979)}]{thinning}%
  \BibitemOpen
  \bibfield  {author} {\bibinfo {author} {\bibfnamefont {P.~A.~W.}\
  \bibnamefont {Lewis}}\ and\ \bibinfo {author} {\bibfnamefont {G.~S.}\
  \bibnamefont {Shedler}},\ }\bibfield  {title} {\bibinfo {title} {Simulation
  of nonhomogeneous poisson process by thinning},\ }\href@noop {} {\bibfield
  {journal} {\bibinfo  {journal} {Naval Res. Logist. Quart.}\ }\textbf
  {\bibinfo {volume} {26}},\ \bibinfo {pages} {403} (\bibinfo {year}
  {1979})}\BibitemShut {NoStop}%
\bibitem [{\citenamefont {Olkin}\ and\ \citenamefont
  {Marshall}(2014)}]{majorization}%
  \BibitemOpen
  \bibfield  {author} {\bibinfo {author} {\bibfnamefont {I.}~\bibnamefont
  {Olkin}}\ and\ \bibinfo {author} {\bibfnamefont {A.}~\bibnamefont
  {Marshall}},\ }\href {https://books.google.fr/books?id=vCPLCQAAQBAJ} {\emph
  {\bibinfo {title} {Inequalities: Theory of Majorization and Its
  Applications}}},\ Mathematics in Science and Engineering\ (\bibinfo
  {publisher} {Academic Press},\ \bibinfo {year} {2014})\BibitemShut {NoStop}%
\bibitem [{\citenamefont {Neves}\ and\ \citenamefont
  {Schonmann}(1991)}]{expEbeta1991}%
  \BibitemOpen
  \bibfield  {author} {\bibinfo {author} {\bibfnamefont {E.}~\bibnamefont
  {Neves}}\ and\ \bibinfo {author} {\bibfnamefont {R.}~\bibnamefont
  {Schonmann}},\ }\bibfield  {title} {\bibinfo {title} {Critical droplets and
  metastability for a glauber dynamics at very low temperatures.},\ }\href
  {https://doi.org/10.1007/BF02431878} {\bibfield  {journal} {\bibinfo
  {journal} {Commun. Math. Phys.}\ }\textbf {\bibinfo {volume} {137}},\
  \bibinfo {pages} {209} (\bibinfo {year} {1991})}\BibitemShut {NoStop}%
\bibitem [{\citenamefont {Rikvold}\ \emph {et~al.}(1994)\citenamefont
  {Rikvold}, \citenamefont {Tomita}, \citenamefont {Miyashita},\ and\
  \citenamefont {Sides}}]{rikvold1994}%
  \BibitemOpen
  \bibfield  {author} {\bibinfo {author} {\bibfnamefont {P.~A.}\ \bibnamefont
  {Rikvold}}, \bibinfo {author} {\bibfnamefont {H.}~\bibnamefont {Tomita}},
  \bibinfo {author} {\bibfnamefont {S.}~\bibnamefont {Miyashita}},\ and\
  \bibinfo {author} {\bibfnamefont {S.~W.}\ \bibnamefont {Sides}},\ }\bibfield
  {title} {\bibinfo {title} {Metastable lifetimes in a kinetic ising model:
  Dependence on field and system size},\ }\href
  {https://doi.org/10.1103/PhysRevE.49.5080} {\bibfield  {journal} {\bibinfo
  {journal} {Phys. Rev. E}\ }\textbf {\bibinfo {volume} {49}},\ \bibinfo
  {pages} {5080} (\bibinfo {year} {1994})}\BibitemShut {NoStop}%
\bibitem [{\citenamefont {Bovier}\ and\ \citenamefont
  {Manzo}(2002)}]{SD-rates}%
  \BibitemOpen
  \bibfield  {author} {\bibinfo {author} {\bibfnamefont {A.}~\bibnamefont
  {Bovier}}\ and\ \bibinfo {author} {\bibfnamefont {F.}~\bibnamefont {Manzo}},\
  }\bibfield  {title} {\bibinfo {title} {Metastability in glauber dynamics in
  the low-temperature limit: Beyond exponential asymptoticsa},\ }\href
  {https://doi.org/10.1023/A:1014586130046} {\bibfield  {journal} {\bibinfo
  {journal} {J. Stat. Phys.}\ }\textbf {\bibinfo {volume} {107}},\ \bibinfo
  {pages} {757} (\bibinfo {year} {2002})},\ \bibinfo {note} {e-print
  cond-mat/0107376}\BibitemShut {NoStop}%
\bibitem [{\citenamefont {Shneidman}\ and\ \citenamefont
  {Nita}(2003)}]{nita2003}%
  \BibitemOpen
  \bibfield  {author} {\bibinfo {author} {\bibfnamefont {V.~A.}\ \bibnamefont
  {Shneidman}}\ and\ \bibinfo {author} {\bibfnamefont {G.~M.}\ \bibnamefont
  {Nita}},\ }\bibfield  {title} {\bibinfo {title} {Nucleation preexponential in
  dynamic ising models at moderately strong fields},\ }\href
  {https://doi.org/10.1103/PhysRevE.68.021605} {\bibfield  {journal} {\bibinfo
  {journal} {Phys. Rev. E}\ }\textbf {\bibinfo {volume} {68}},\ \bibinfo
  {pages} {021605} (\bibinfo {year} {2003})}\BibitemShut {NoStop}%
\bibitem [{\citenamefont {Novotny}(2002)}]{novotny2002-3D}%
  \BibitemOpen
  \bibfield  {author} {\bibinfo {author} {\bibfnamefont {M.}~\bibnamefont
  {Novotny}},\ }\bibfield  {title} {\bibinfo {title} {Low-temperature long-time
  simulations of ising ferromagnets using the monte carlo with absorbing markov
  chains method},\ }\href
  {https://doi.org/https://doi.org/10.1016/S0010-4655(02)00369-7} {\bibfield
  {journal} {\bibinfo  {journal} {Computer Physics Communications}\ }\textbf
  {\bibinfo {volume} {147}},\ \bibinfo {pages} {659} (\bibinfo {year}
  {2002})},\ \bibinfo {note} {proceedings of the Europhysics Conference on
  Computational Physics Computational Modeling and Simulation of Complex
  Systems}\BibitemShut {NoStop}%
\bibitem [{\citenamefont {Sides}\ \emph
  {et~al.}(1998{\natexlab{b}})\citenamefont {Sides}, \citenamefont {Rikvold},\
  and\ \citenamefont {Novotny}}]{droplet-theory98}%
  \BibitemOpen
  \bibfield  {author} {\bibinfo {author} {\bibfnamefont {S.~W.}\ \bibnamefont
  {Sides}}, \bibinfo {author} {\bibfnamefont {P.~A.}\ \bibnamefont {Rikvold}},\
  and\ \bibinfo {author} {\bibfnamefont {M.~A.}\ \bibnamefont {Novotny}},\
  }\bibfield  {title} {\bibinfo {title} {{Hysteresis loop areas in kinetic
  Ising models: Effects of the switching mechanism}},\ }\href
  {https://doi.org/10.1063/1.367600} {\bibfield  {journal} {\bibinfo  {journal}
  {J. Appl. Phys.}\ }\textbf {\bibinfo {volume} {83}},\ \bibinfo {pages} {6494}
  (\bibinfo {year} {1998}{\natexlab{b}})},\ \bibinfo {note} {e-print
  cond-mat/9710244}\BibitemShut {NoStop}%
\bibitem [{\citenamefont {Glauber}(1963)}]{glauber}%
  \BibitemOpen
  \bibfield  {author} {\bibinfo {author} {\bibfnamefont {R.~J.}\ \bibnamefont
  {Glauber}},\ }\bibfield  {title} {\bibinfo {title} {Time-dependent statistics
  of the {I}sing model},\ }\href {https://doi.org/10.1063/1.1703954} {\bibfield
   {journal} {\bibinfo  {journal} {J. Math. Phys.}\ }\textbf {\bibinfo {volume}
  {4}},\ \bibinfo {pages} {294} (\bibinfo {year} {1963})}\BibitemShut {NoStop}%
\bibitem [{\citenamefont {Vatansever}\ \emph {et~al.}(2024)\citenamefont
  {Vatansever}, \citenamefont {Vatansever}, \citenamefont {Berger},
  \citenamefont {Vasilopoulos},\ and\ \citenamefont {Fytas}}]{2componentH}%
  \BibitemOpen
  \bibfield  {author} {\bibinfo {author} {\bibfnamefont {Z.~D.}\ \bibnamefont
  {Vatansever}}, \bibinfo {author} {\bibfnamefont {E.}~\bibnamefont
  {Vatansever}}, \bibinfo {author} {\bibfnamefont {A.}~\bibnamefont {Berger}},
  \bibinfo {author} {\bibfnamefont {A.}~\bibnamefont {Vasilopoulos}},\ and\
  \bibinfo {author} {\bibfnamefont {N.~G.}\ \bibnamefont {Fytas}},\ }\bibfield
  {title} {\bibinfo {title} {Monte carlo study of the two-dimensional kinetic
  {I}sing model under a nonantisymmetric magnetic field},\ }\href
  {https://doi.org/10.1103/PhysRevE.110.064155} {\bibfield  {journal} {\bibinfo
   {journal} {Phys. Rev. E}\ }\textbf {\bibinfo {volume} {110}},\ \bibinfo
  {pages} {064155} (\bibinfo {year} {2024})}\BibitemShut {NoStop}%
\bibitem [{\citenamefont {Park}\ \emph {et~al.}(2004)\citenamefont {Park},
  \citenamefont {Rikvold}, \citenamefont {Buend{\'{\i}}a},\ and\ \citenamefont
  {Novotny}}]{soft-dynamics}%
  \BibitemOpen
  \bibfield  {author} {\bibinfo {author} {\bibfnamefont {K.}~\bibnamefont
  {Park}}, \bibinfo {author} {\bibfnamefont {P.~A.}\ \bibnamefont {Rikvold}},
  \bibinfo {author} {\bibfnamefont {G.~M.}\ \bibnamefont {Buend{\'{\i}}a}},\
  and\ \bibinfo {author} {\bibfnamefont {M.~A.}\ \bibnamefont {Novotny}},\
  }\bibfield  {title} {\bibinfo {title} {Low-temperature nucleation in a
  kinetic {I}sing model with soft stochastic dynamics},\ }\href@noop {}
  {\bibfield  {journal} {\bibinfo  {journal} {Phys. Rev. Lett.}\ }\textbf
  {\bibinfo {volume} {92}},\ \bibinfo {pages} {015701} (\bibinfo {year}
  {2004})}\BibitemShut {NoStop}%
\bibitem [{\citenamefont {Buend\'{\i}a}\ and\ \citenamefont
  {Rikvold}(2017)}]{DPT2017}%
  \BibitemOpen
  \bibfield  {author} {\bibinfo {author} {\bibfnamefont {G.~M.}\ \bibnamefont
  {Buend\'{\i}a}}\ and\ \bibinfo {author} {\bibfnamefont {P.~A.}\ \bibnamefont
  {Rikvold}},\ }\bibfield  {title} {\bibinfo {title} {Fluctuations in a model
  ferromagnetic film driven by a slowly oscillating field with a constant
  bias},\ }\href {https://doi.org/10.1103/PhysRevB.96.134306} {\bibfield
  {journal} {\bibinfo  {journal} {Phys. Rev. B}\ }\textbf {\bibinfo {volume}
  {96}},\ \bibinfo {pages} {134306} (\bibinfo {year} {2017})}\BibitemShut
  {NoStop}%
\bibitem [{\citenamefont {Suen}\ \emph {et~al.}(1999)\citenamefont {Suen},
  \citenamefont {Lee}, \citenamefont {Teeter},\ and\ \citenamefont
  {Erskine}}]{hyst-thin-films}%
  \BibitemOpen
  \bibfield  {author} {\bibinfo {author} {\bibfnamefont {J.-S.}\ \bibnamefont
  {Suen}}, \bibinfo {author} {\bibfnamefont {M.~H.}\ \bibnamefont {Lee}},
  \bibinfo {author} {\bibfnamefont {G.}~\bibnamefont {Teeter}},\ and\ \bibinfo
  {author} {\bibfnamefont {J.~L.}\ \bibnamefont {Erskine}},\ }\bibfield
  {title} {\bibinfo {title} {Magnetic hysteresis dynamics of thin co films on
  cu(001)},\ }\href {https://doi.org/10.1103/PhysRevB.59.4249} {\bibfield
  {journal} {\bibinfo  {journal} {Phys. Rev. B}\ }\textbf {\bibinfo {volume}
  {59}},\ \bibinfo {pages} {4249} (\bibinfo {year} {1999})}\BibitemShut
  {NoStop}%
\bibitem [{\citenamefont {Suen}\ and\ \citenamefont {Erskine}(1997)}]{Fe_on_W}%
  \BibitemOpen
  \bibfield  {author} {\bibinfo {author} {\bibfnamefont {J.-S.}\ \bibnamefont
  {Suen}}\ and\ \bibinfo {author} {\bibfnamefont {J.~L.}\ \bibnamefont
  {Erskine}},\ }\bibfield  {title} {\bibinfo {title} {Magnetic hysteresis
  dynamics: Thin $\mathit{p}(1\ifmmode\times\else\texttimes\fi{}1)$ fe films on
  flat and stepped w(110)},\ }\href
  {https://doi.org/10.1103/PhysRevLett.78.3567} {\bibfield  {journal} {\bibinfo
   {journal} {Phys. Rev. Lett.}\ }\textbf {\bibinfo {volume} {78}},\ \bibinfo
  {pages} {3567} (\bibinfo {year} {1997})}\BibitemShut {NoStop}%
\bibitem [{\citenamefont {Brockett}(1970)}]{brockett1970}%
  \BibitemOpen
  \bibfield  {author} {\bibinfo {author} {\bibfnamefont {R.}~\bibnamefont
  {Brockett}},\ }\href@noop {} {\emph {\bibinfo {title} {Finite Dimensional
  Linear Systems}}}\ (\bibinfo  {publisher} {John Wiley \& Sons, Inc.},\
  \bibinfo {address} {New York},\ \bibinfo {year} {1970})\BibitemShut {NoStop}%
\bibitem [{\citenamefont {Shneidman}(2003)}]{Shneidman_2003}%
  \BibitemOpen
  \bibfield  {author} {\bibinfo {author} {\bibfnamefont {V.}~\bibnamefont
  {Shneidman}},\ }\bibfield  {title} {\bibinfo {title} {On the lowest energy
  nucleation path in a supersaturated lattice gas.},\ }\href
  {https://doi.org/10.1023/A:1023687822656} {\bibfield  {journal} {\bibinfo
  {journal} {J. Stat. Phys.}\ }\textbf {\bibinfo {volume} {112}},\ \bibinfo
  {pages} {293} (\bibinfo {year} {2003})}\BibitemShut {NoStop}%
\bibitem [{\citenamefont {Shneidman}(2005)}]{Shneidman_2005}%
  \BibitemOpen
  \bibfield  {author} {\bibinfo {author} {\bibfnamefont {V.~A.}\ \bibnamefont
  {Shneidman}},\ }\bibfield  {title} {\bibinfo {title} {Branching of nucleation
  paths in a metastable lattice gas with metropolis dynamics},\ }\href
  {https://doi.org/10.1088/1367-2630/7/1/012} {\bibfield  {journal} {\bibinfo
  {journal} {New J. Phys.}\ }\textbf {\bibinfo {volume} {7}},\ \bibinfo {pages}
  {12} (\bibinfo {year} {2005})}\BibitemShut {NoStop}%
\bibitem [{\citenamefont {Korniss}\ \emph {et~al.}(2002)\citenamefont
  {Korniss}, \citenamefont {Rikvold},\ and\ \citenamefont {Novotny}}]{2dIsing}%
  \BibitemOpen
  \bibfield  {author} {\bibinfo {author} {\bibfnamefont {G.}~\bibnamefont
  {Korniss}}, \bibinfo {author} {\bibfnamefont {P.~A.}\ \bibnamefont
  {Rikvold}},\ and\ \bibinfo {author} {\bibfnamefont {M.~A.}\ \bibnamefont
  {Novotny}},\ }\bibfield  {title} {\bibinfo {title} {Absence of first-order
  transition and tricritical point in the dynamic phase diagram of a spatially
  extended bistable system in an oscillating field},\ }\href
  {https://doi.org/10.1103/PhysRevE.66.056127} {\bibfield  {journal} {\bibinfo
  {journal} {Phys. Rev. E}\ }\textbf {\bibinfo {volume} {66}},\ \bibinfo
  {pages} {056127} (\bibinfo {year} {2002})}\BibitemShut {NoStop}%
\bibitem [{\citenamefont {Shneidman}\ and\ \citenamefont
  {Nita}(2002)}]{nita02}%
  \BibitemOpen
  \bibfield  {author} {\bibinfo {author} {\bibfnamefont {V.~A.}\ \bibnamefont
  {Shneidman}}\ and\ \bibinfo {author} {\bibfnamefont {G.~M.}\ \bibnamefont
  {Nita}},\ }\bibfield  {title} {\bibinfo {title} {Modulation of the nucleation
  rate preexponential in a low-temperature ising system},\ }\href
  {https://doi.org/10.1103/PhysRevLett.89.025701} {\bibfield  {journal}
  {\bibinfo  {journal} {Phys. Rev. Lett.}\ }\textbf {\bibinfo {volume} {89}},\
  \bibinfo {pages} {025701} (\bibinfo {year} {2002})}\BibitemShut {NoStop}%
\bibitem [{\citenamefont {Novotny}(2003)}]{prefactors2003}%
  \BibitemOpen
  \bibfield  {author} {\bibinfo {author} {\bibfnamefont {M.~A.}\ \bibnamefont
  {Novotny}},\ }\bibfield  {title} {\bibinfo {title} {Low-temperature
  metastability of ising models: Prefactors, divergences, and
  discontinuities},\ }in\ \href {https://doi.org/10.1007/978-3-642-55522-0_2}
  {\emph {\bibinfo {booktitle} {Computer Simulation Studies in Condensed-Matter
  Physics XV}}},\ \bibinfo {editor} {edited by\ \bibinfo {editor}
  {\bibfnamefont {D.~P.}\ \bibnamefont {Landau}}, \bibinfo {editor}
  {\bibfnamefont {S.~P.}\ \bibnamefont {Lewis}},\ and\ \bibinfo {editor}
  {\bibfnamefont {H.-B.}\ \bibnamefont {Sch{\"u}ttler}}}\ (\bibinfo
  {publisher} {Springer Berlin Heidelberg},\ \bibinfo {address} {Berlin,
  Heidelberg},\ \bibinfo {year} {2003})\ pp.\ \bibinfo {pages}
  {7--19}\BibitemShut {NoStop}%
\bibitem [{\citenamefont {Bender}\ and\ \citenamefont
  {Orszag}(1999)}]{bender1999advanced}%
  \BibitemOpen
  \bibfield  {author} {\bibinfo {author} {\bibfnamefont {C.}~\bibnamefont
  {Bender}}\ and\ \bibinfo {author} {\bibfnamefont {S.}~\bibnamefont
  {Orszag}},\ }\href {https://books.google.fr/books?id=-yQXwhE6iWMC} {\emph
  {\bibinfo {title} {Advanced Mathematical Methods for Scientists and Engineers
  I: Asymptotic Methods and Perturbation Theory}}},\ Advanced Mathematical
  Methods for Scientists and Engineers\ (\bibinfo  {publisher} {Springer},\
  \bibinfo {year} {1999})\BibitemShut {NoStop}%
\end{thebibliography}
\end{document}